\documentclass[prl,aps,reprint,superscriptaddress]{revtex4-1}
\usepackage{amsmath}
\usepackage{graphics}
\usepackage{graphicx}
\usepackage{epsfig}
\usepackage{hyperref}
\usepackage{color}
\usepackage{amsfonts}
\usepackage[normalem]{ulem}

\usepackage{textcomp}

\newcommand{\beq}{\begin{equation}}
\newcommand{\eeq}{\end{equation}}
\newcommand{\beqar}{\begin{eqnarray}}
\newcommand{\eeqar}{\end{eqnarray}}
\newcommand{\bcen}{\begin{center}}
\newcommand{\ecen}{\end{center}}

\newcommand{\blue}[1]{{\color{blue} #1}}

\begin{document}

\title{RIXS Reveals Hidden Local Transitions of the Aqueous OH Radical} %\\

\author{L. Kjellsson}
\affiliation{Department of Physics and Astronomy, Uppsala University, Box 516, S-751 20 Uppsala, Sweden}

\author{K. Nanda}
\affiliation{Department of Chemistry, University of Southern California, United States}

\author{J.-E. Rubensson}
\affiliation{Department of Physics and Astronomy, Uppsala University, Box 516, S-751 20 Uppsala, Sweden}

\author{G. Doumy}
\affiliation{Chemical Sciences and Engineering Division, Argonne National Laboratory, United States}

\author{S. H. Southworth}
\affiliation{Chemical Sciences and Engineering Division, Argonne National Laboratory, United States}

\author{P. J. Ho}
\affiliation{Chemical Sciences and Engineering Division, Argonne National Laboratory, United States}

\author{A. M. March}
\affiliation{Chemical Sciences and Engineering Division, Argonne National Laboratory, United States}

\author{A. Al Haddad}
\affiliation{Chemical Sciences and Engineering Division, Argonne National Laboratory, United States}

\author{Y. Kumagai}
\affiliation{Chemical Sciences and Engineering Division, Argonne National Laboratory, United States}

\author{M.-F. Tu}
\affiliation{Chemical Sciences and Engineering Division, Argonne National Laboratory, United States}

\author{R. Schaller}
\affiliation{Center for Nanoscale Materials, Argonne National Laboratory, United States}
\affiliation{Department of Chemistry, Northwestern University, United States}

\author{T. Debnath}
\affiliation{Division of Chemistry and Biological Chemistry, Nanyang Technological University, Singapore}

\author{M. S. Bin Mohd Yusof}
\affiliation{Division of Chemistry and Biological Chemistry, Nanyang Technological University, Singapore}

\author{C. Arnold}
\affiliation{Center for Free-Electron Laser Science, DESY, Germany}
\affiliation{Department of Physics, Universität Hamburg, Germany}
\affiliation{Hamburg Centre for Ultrafast Imaging, Hamburg, Germany}

\author{W. F. Schlotter}
\affiliation{LCLS, SLAC National Accelerator Laboratory, United States}

\author{S. Moeller}
\affiliation{LCLS, SLAC National Accelerator Laboratory, United States}

\author{G. Coslovich}
\affiliation{LCLS, SLAC National Accelerator Laboratory, United States}

\author{J. D. Koralek}
\affiliation{LCLS, SLAC National Accelerator Laboratory, United States}

\author{M. P. Minitti}
\affiliation{LCLS, SLAC National Accelerator Laboratory, United States}

\author{M. L. Vidal}
\affiliation{DTU Chemistry - Department of Chemistry, Technical University of Denmark, DK-2800, Kongens Lyngby, Denmark}

\author{M. Simon}
\affiliation{Sorbonne Université and CNRS, Laboratoire de Chimie Physique-Matière et Rayonnement, France}

\author{R. Santra}
\affiliation{Center for Free-Electron Laser Science, DESY, Germany}
\affiliation{Department of Physics, Universität Hamburg, Germany}
\affiliation{Hamburg Centre for Ultrafast Imaging, Hamburg, Germany}

\author{Z.-H. Loh}
\affiliation{Division of Chemistry and Biological Chemistry, Nanyang Technological University, Singapore}

\author{S. Coriani}
\affiliation{DTU Chemistry - Department of Chemistry, Technical University of Denmark, DK-2800, Kongens Lyngby, Denmark}

\author{A. I. Krylov}
\email{krylov@usc.edu}
\affiliation{Department of Chemistry, University of Southern California, United States}

\author{L. Young}
\email{young@anl.gov}
\affiliation{Chemical Sciences and Engineering Division, Argonne National Laboratory, United States}
\affiliation{Department of Physics and James Franck Institute, The University of Chicago, United States}

\date{\today }

\begin{abstract}
Resonant inelastic x-ray scattering (RIXS)  provides remarkable opportunities to interrogate ultrafast dynamics in liquids.  Here we use RIXS to study the fundamentally and practically important hydroxyl radical in liquid water, OH$(aq)$. Impulsive ionization of pure liquid water produced a short-lived population of OH$(aq)$, which was probed using femtosecond x-rays from an x-ray free-electron laser.  We find that RIXS reveals localized electronic transitions that are masked in the ultraviolet absorption spectrum by strong charge-transfer transitions\textemdash thus providing a means to investigate the evolving electronic structure and reactivity of the hydroxyl radical in aqueous and heterogeneous environments. First-principles calculations provide interpretation of the main spectral features.

\end{abstract}

\maketitle

%\section{Introduction}

The hydroxyl radical (OH) is of major importance for atmospheric, astrochemical, biological, industrial, and environmental research. In the gas phase, OH is the primary oxidizing agent that rids the atmosphere of volatile organic compounds and other pollutants \cite{Isaksen-2011-Science}. It is a key tracer describing the evolution and thermodynamics of interstellar clouds \cite{Rank-1971-Science}. Despite its reactive nature stemming from an open electronic shell, the gas\blue{-}phase absorption spectrum of OH has been fully characterized in the microwave \cite{Robinson-1967-ARAA}, infrared, optical/ultraviolet (UV) \cite{huber1979}, and, more recently, the x-ray \cite {Stranges-2002-JCP} spectral ranges. Beyond purely gas-phase processes, these OH fingerprints also characterize heterogeneous processes such as the generation of reactive oxygen species from photocatalysis \cite{Nosaka-2014-CR}.

\begin{figure}[!htbp!]
\centering
\includegraphics[width=8.7cm]{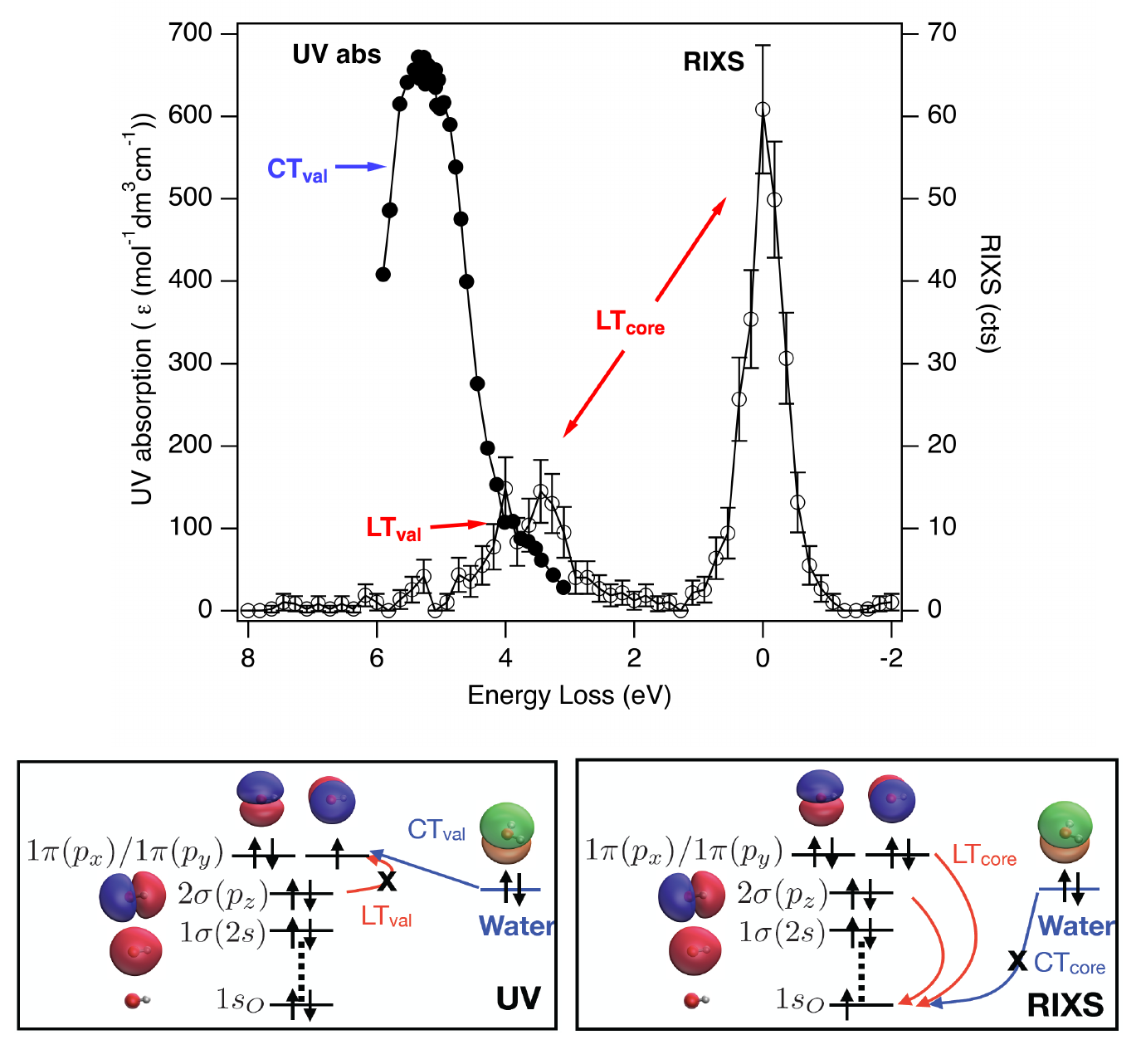}
\caption{UV absorption and RIXS energy-loss spectra of OH$(aq)$ (top) and molecular orbital diagrams (bottom). The electronic configuration of OH is $1s_O^2 1\sigma^2 2\sigma^2 1\pi^3$.
The lowest valence transition ($2\sigma\rightarrow 1\pi$, marked as LT$_{val}$ for local transition) has low oscillator strength due to its $p_z\rightarrow p_x/p_y$ character; the UV spectrum is dominated by charge-transfer transition (CT$_{val}$) from nearby waters. In resonantly excited OH ($1s_O^1 1\sigma^2 2\sigma^2 1\pi^4$), both $1\pi \rightarrow 1s$ and $2\sigma\rightarrow 1s$ transitions are bright, giving rise to the elastic peak (zero energy loss) and a feature at $\sim$4 eV. The gap between the two RIXS peaks corresponds to the $2\sigma\rightarrow 1\pi$ energy. %The CT transitions (to $1s_O$ of OH) from nearby waters are suppressed due to a compact shape of the 1$s_O$ orbital, resulting in poor overlap. 
\protect\label{fig:SpectraAndMOs}}
\end{figure}

Spectroscopic characterization of the hydroxyl radical in the condensed phase is more challenging, owing to its extreme reactivity and short lifetime.
Of particular interest is the characterization of OH in aqueous environments, which impacts radiation biology \cite{Alizadeh-2012-CR} and chemistry \cite{Garrett2005}. Given its unpaired spin, electron spin resonance techniques are a natural choice to detect the presence of OH, either directly or through spin traps, but only microsecond timescales are accessible \cite{Nosaka-2014-CR}.  For faster timescales, desired for tracking reaction dynamics, one may consider UV spectroscopy.  The UV spectrum of solvated OH obtained via pulsed radiolysis of water \cite{Czapski1993} is reproduced in Fig. \ref{fig:SpectraAndMOs}. It is dominated by a strong feature at 230 nm (5.4 eV), whereas the dominant gas-phase absorption at 309 nm (4 eV), due to valence excitation from the ground (X) to the lowest excited electronic state (A), is barely visible. 
On the basis of electronic structure calculations \cite{Hamad2002,Chipman2008,Chipman2011,Cordoniu2013}, this dominant spectral feature of OH($aq$) 
was attributed to charge-transfer (CT) transitions from the lone pair of nearby waters, filling the hole in the OH $1\pi$ orbital (Fig.  \ref{fig:SpectraAndMOs}, bottom left).

Resonant inelastic x-ray scattering (RIXS) delivers atomic-site specific information about the local electronic structure and dynamics in condensed phase. The application of soft x-ray RIXS to liquids \cite{Guo2002} has generated considerable attention; improvements in sensitivity \cite{Fuchs2008} and energy resolution \cite{Hennies2010} continue to open new perspectives on fundamental liquid-phase interactions \cite{Sun2011, Harada2013, VazdaCruz2019}.  Recently, the combination of RIXS, liquid microjets, and x-ray free-electron lasers enabled time-resolved measurements of electronic structure of transient species in solutions \cite{Wernet2015, Jay2018}.

Here we report the RIXS spectrum for the short-lived OH radical in water (Fig. \ref{fig:SpectraAndMOs}).  In sharp contrast to the UV spectrum, the RIXS spectrum of OH$(aq)$ features two peaks corresponding to transitions between the OH orbitals (Fig. \ref{fig:SpectraAndMOs}, bottom right). The energy difference between the elastic and inelastic peaks corresponds to the X$\rightarrow$A transition,  which, in turn, roughly equals the energy gap between the $2\sigma$ and $1\pi$ orbitals.
Thus, RIXS reveals intrinsic local electronic structure of solvated OH, which is obscured in the UV region by CT transitions. 
In contrast to CT transitions, which are characteristic of the solvent and its local structure, local transitions (LT) are fingerprints of the solute and can, therefore, be used to track reactive hydroxyl radicals in various complex and heterogeneous environments.
The CT transitions in the RIXS spectrum are suppressed because the compact shape of the core $1s_O$ orbital results in poor overlap with the lone pairs of neighboring waters. We confirm this by \emph{ab initio} RIXS calculations using a new electronic structure method \cite{Vidal:CVSEOM:18,Nanda:RIXS:19} based on the equation-of-motion coupled-cluster (EOM-CC) theory. 
These calculations also reproduce the relative RIXS line intensities, positions, and widths for the elastic and inelastic peaks of OH$(aq)$ and OH$^-(aq)$.  

The OH$(aq)$ spectra are also compared to the x-ray emission spectra (XES) of liquid water, where the role of hydrogen bonding and ultrafast dynamics has long been debated  \cite{Fuchs2008,Tokushima2008,Harada2013,Pietzsch2015PRL,Fransson2016,Yamazoe2019,Niskanen2019}. If core-ionized water dissociates prior to core-hole decay, a core-excited OH$(aq)$ is formed in the $1s_O^1 1\sigma^2 2\sigma^2 1\pi^4$ intermediate RIXS state (Fig.\ref{fig:SpectraAndMOs}). The lower-energy component of the water XES doublet has been attributed to this ultrafast dissociation \cite{Fuchs2008} and our measurement of the position of the OH$(aq)$ RIXS resonance directly provides relevant information that previously was indirectly deduced \cite{Yamazoe2019}.     

We create the transient hydroxyl radical using strong-field ionization in pure liquid water \cite{Loh2020Science}. The laser-induced ionization initially forms a water cation (H$_2$O$^+$), which undergoes ultrafast proton transfer with a neighboring water molecule on the sub-100-fs timescale \cite{Kamarchik:10:Water_PES} forming the hydroxyl radical and hydronium ion (H$_3$O$^+$). In the time window between proton transfer ($\sim$100 fs) and geminate recombination, the ionized liquid water sample contains hydroxyl radicals.  Because the $1s_O\rightarrow 1\pi$ transition in OH$(aq)$ occurs cleanly in the ``water window'', i.e., below the liquid water absorption edge, its kinetics can be readily probed via transient absorption \cite{Loh2020Science} and, simultaneously, via RIXS. Importantly, the latter allows investigation of local valence transitions, which are  chemically most relevant.

%\section{Experimental} 
% SXR Setup

% Scans combined for rixsmap: 53,55,58,63
% There was a shift change between 58 and 63

Briefly, optical-pump x-ray-probe RIXS was performed using the Soft X-ray Research (SXR) instrument \cite{schlotter2012} at the Linac Coherent Light Source (LCLS) at SLAC National Accelerator Laboratory. Monochromatized x-ray pulses were scanned from 518 to 542 eV ($\sim10\mu$J, $\sim40$fs, 200 meV bandwidth). Three photon detection channels were simultaneously recorded: transmission, total fluorescence, and dispersed emission. %The monochromator exit slit was 130 {\textmu}m with a calculated bandwidth of 200 meV FWHM.  
A detailed description of the performance of the optical laser, water jet, and x-ray monochromator calibration, shot-by-shot normalization procedures can be found in \cite{Loh2020Science}. Here we describe additionally the x-ray emission spectrometer and experimental geometries for RIXS measurements of OH$(aq)$.

% XES Spectrometer

X-ray emission was collected perpendicular to the incoming x-ray beam and along the x-ray polarization axis using a variable-line-spacing grating-based spectrometer \cite{chuang2017modularVLSSPEC}. A CCD camera located at the exit plane of the spectrometer recorded images on a shot-by-shot basis. %XFEL pulses with very low intensity, as measured by the I$_0$ detector, were not considered in the analysis. After filtering from a total of 370000 pulses, we were left with 15000 counts of the XES.

\begin{figure*}[!hbt]
\centering
\includegraphics{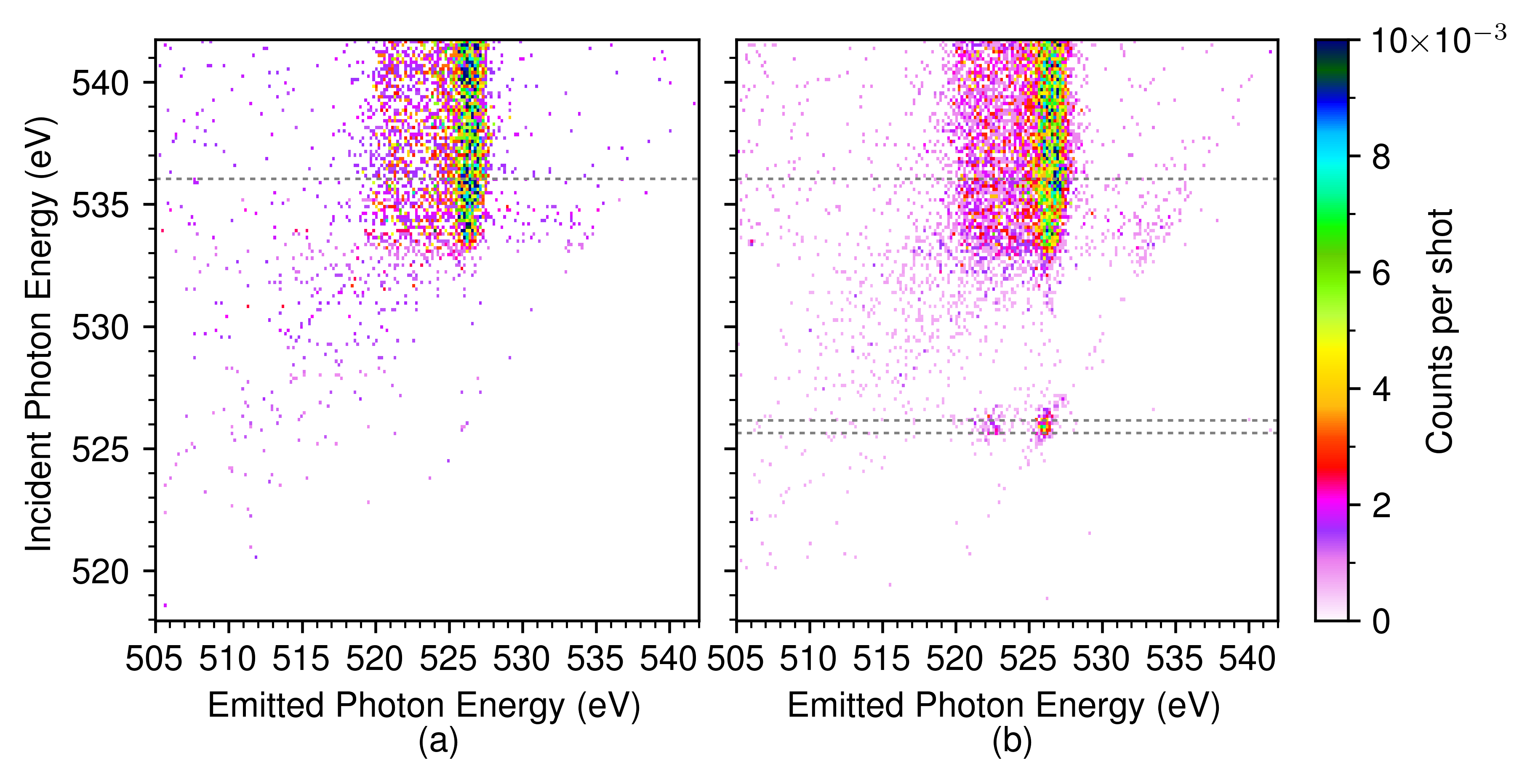} % More contrast needed?

\caption{RIXS maps of ionized liquid water. (a) before valence ionization, (b) after valence ionization, integrated for time delays between 200-1400 fs. The intensity of the maps is the number of counts on the emission spectrometer (15054) normalized by the number of XFEL pulses (365846). Before ionization, a threshold for emission is seen near 534 eV incident photon energy. %and a resonance with a pre-peak at 535 eV. 
After ionization, a resonant feature appears at 526 eV due to the transient OH$(aq)$ radical. The two energy windows used to create the spectra shown in Fig. \ref{fig:XES-combo} are marked: 536-542 eV for bulk water and 525.70-526.12 eV for OH$(aq)$.  
}
\label{fig:rixsmap}
\end{figure*}

% Incoming energy calibration
 The energy dispersion and absolute energy of the incoming monochromatized radiation were previously calibrated  \cite{Loh2020Science, nagasaka2010}.
% XES Energy Calibration
The dispersion of the emission spectrometer was determined by fitting a first degree polynomial to the elastic line visible in Fig. \ref{fig:rixsmap}(b).

%%-------------------------------------------%%%%%%%%%%%%%%%%%%%%%%%%%%%%%%%%%
To gain insight into the nature of main spectral features, we carried out electronic structure calculations using EOM-CC \cite{Krylov:EOMRev:07} with single and double excitations (EOM-CCSD), augmented by core-valence separation (CVS) \cite{Cederbaum:CVSorig:80} to enable access to core-level states \cite{Coriani:CVSEOM:15,Vidal:CVSEOM:18}. 
As a multistate method, EOM-CC treats different valence and core-level states on an equal footing and is particularly well suited for modeling molecular properties, including non-linear  properties  \cite{Krylov:EOMRev:07,Helgaker:Resp_theory_rev:2012,Nanda:2PA:14,Nanda:2PAEFP:2018}.
To account for solvent effects, the spectral calculations of  OH$(aq)$/OH$^-(aq)$ were carried out within the QM/MM (quantum mechanics-molecular mechanics) scheme with water molecules described by classical force field and OH/OH$^-$ described by EOM-CCSD by using snapshots from equilibrium {\em ab initio} molecular dynamics simulations. 

%\setcounter{bottomnumber}{2}
%\renewcommand{\topfraction}{0.25}
%\begin{figure*}[!htbp!]

The electronic factors entering RIXS cross sections 
are  the   RIXS transition  moments   given  by  the   following  
Kramers-Heisenberg-Dirac (KHD) expression \cite{Gelmukhanov:RIXS:99}:
\begin{eqnarray}
  \begin{split}
%\left(\omega_{i,x}, - \omega_{e,y}\right)
    &M_{fg}^{xy} =  \nonumber 
    -\sum_{n} \left(\frac{\langle f |\mu^{y}| n\rangle\langle n |\mu^{x}| g\rangle}{\Omega_{ng} - \omega_{i,x} - i\epsilon_n} 
     +~ \frac{\langle f|\mu^{x}| n\rangle\langle n|\mu^{y}|g\rangle}{\Omega_{ng} + \omega_{o,y} + i\epsilon_n}\right),
\label{rixs-khd}
  \end{split}
\end{eqnarray}
where $g$ and $f$ denote the initial and final electronic states (i.e., ground and valence excited state of OH), $\omega_i$/$\omega_o$ are the incoming/outgoing photon frequencies, and 
 the sum runs over all electronic states; 
$\Omega_{ng}  = E_n  -  E_g$  is the  energy
difference between states $n$ and $g$, and $i\epsilon_n$ is the imaginary
inverse lifetime  parameter for  state $n$. 
In the present experiment, the dominant contribution to the 
RIXS cross section comes from the term corresponding to the $1s_O^1\ldots1\pi^4$ state, which is resonant 
with excitation frequency of 526 eV, such that the spectra can be qualitatively understood within a three-states model. 
Within the  EOM-CC framework,  the KHD  expression is  evaluated using 
EOM-CC energies  and wave functions.
Rather than arbitrarily truncating the sum over states, we replace all $\epsilon_n$s with    a    phenomenological    damping    factor    $\epsilon$ and use damped response theory to convert the KHD expression into a numerically tractable closed form \cite{Nanda:2PA:14,Nanda:RIXS:19,Faber2019JCTC}.
Robust convergence of the auxiliary response equations is achieved by  using CVS within the damped response domain \cite{Nanda:RIXS:19}. The resulting method combines rigorous treatment of RIXS cross sections and high-level description of electron correlation.
%In   all EOM-CCSD   calculations,   we    used   the 6-311(2+,+)G(2df,p) basis  with uncontracted oxygen core, denoted as uC-6-311(2+,+)G(2df,p),  which has been      shown to adequately describe core-level states \cite{Nanda:RIXS:19}.
To describe vibrational structure in the RIXS spectrum, we computed Franck-Condon factors (FCFs) using three-states model (as was done in Ref. \cite{Sun2011}) and harmonic approximation. To quantify relative strengths of local and CT transitions, we also carried out calculations on model water-OH structures. All calculations were performed using the Q-Chem electronic structure package \cite{qchem_feature}. Full details of computational protocols are given in the SI.

%%%%%%%%%%%%%%%%%%%%%%%%%%%%% Results and discussion %%%%%%%%%%%%%%%%%%%%%%%%%
%\section{Results and Discussion}

Theoretical  estimates of key structural parameters  of the isolated OH given in Table  \ref{tbl:OHgas} agree well with experimental  values \cite{huber1979,Stranges-2002-JCP}. 
Theory overestimates the energy of the valence transition by 0.08 eV and underestimates the energy of the core-excited state by 0.7 eV; these differences are within the error bars of the method \cite{Vidal:CVSEOM:18}.  
The variations in bond lengths and frequencies among different states are consistent with the molecular orbital picture of the electronic states.
%We note that the valence excited state has considerably larger dipole moment than the ground state ($\sim$14\% increase), suggesting a larger inhomogeneous broadening of the energy-loss peak in the RIXS spectrum\cite{Sun2011}.  
The structural differences between the states give rise to a vibrational progression in the x-ray absorption spectrum; the computed FCFs are in excellent agreement with the experimental ones (see SI). 

%\onecolumngrid
\tabcolsep 3.0pt
\begin{table}[!htbp!]
  \caption{Key structural parameters of isolated OH radical.
        \protect\label{tbl:OHgas}}
  \begin{tabular}{llccccl}
\hline    
    State    &   Character          &   $T_e$, eV     & $r_e$, \AA & $\omega_e$, eV & $\mu$, a.u. & \\
\hline 
X($^2\Pi_1$) &  $1s^2 1\sigma^2 2\sigma^2 1\pi^3$            &  0.000           & 0.972       & 0.468   & 0.701  & $^a$ \\
             &             &  0.000           & 0.970      & 0.463   &        & $^b$ \\ 
\hline
core $^2\Sigma^+$ & 1s$^{-1} 1\sigma^2 2\sigma^2  \pi^4$ & 525.1  & 0.916   & 0.543   & 0.814 & $^a$ \\  
                  &                  & 525.8  & 0.915   & 0.533   &       & $^c$ \\ 
\hline
A($^2\Sigma^+$) &   $1s^2 1\sigma^2 2\sigma^1 1\pi^4$   &    4.128            & 1.014  & 0.398 & 0.801 & $^a$ \\  

                &             &   4.052             & 1.012  & 0.394   &     & $^b$ \\
\hline
  \end{tabular}

  $^a$ Theory, this work. Energies ($T_e$) and dipole moments ($\mu$): (cvs)-EOM-EE-CCSD, computed at the experimental geometries;
  $r_e$ and $\omega_e$:  (cvs)-EOM-IP-CCSD. Basis set:  uC-6-311(2+,+)G(2df,p).
$^b$ Expt. Ref. \citenum{huber1979}.
 $^c$ Expt. Ref. \citenum{Stranges-2002-JCP}.
\end{table}

%\twocolumngrid

\begin{figure}[!htbp!]
%\begin{figure}[h]
\centering
\includegraphics[width=8cm] {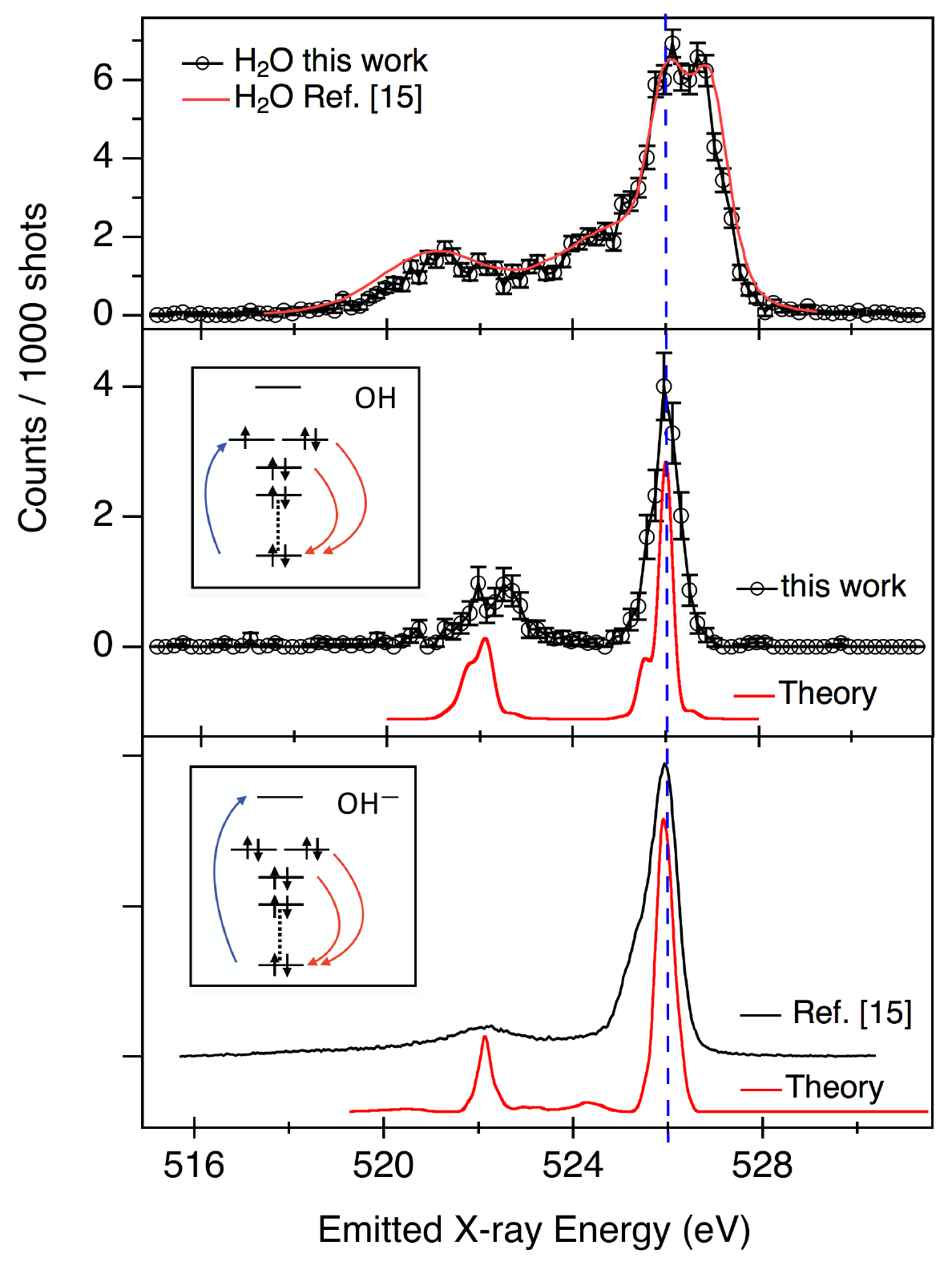}
\caption{Comparison of RIXS from OH and OH$^-$. Top: Experimental XES from water for excitation energies between 536-542 eV compared to XES of water excited at 550.1 eV \cite{Fuchs2008}. Middle: RIXS from the 526-eV resonance of OH$(aq)$ compared with theory (shifted +1.0 eV).  Bottom: RIXS from OH$^-$ excited at 533.5 eV and compared with theory (shifted -0.2 eV). Ref. \cite{Fuchs2008} data shifted -0.7 eV. Insets show molecular orbital diagrams for the RIXS transitions in OH and OH$^-$. 
}
\label{fig:XES-combo}
\end{figure}

% Overview of the result, threshold for emission at 534eV
%%%% ------------------------------------------------------------------------%%%%
Fig. \ref{fig:rixsmap} shows the RIXS maps before and after the ionization pulse.   The RIXS map prior to ionization, Fig. \ref{fig:rixsmap}(a), is in agreement with earlier measurements  \cite{Fuchs2008,Tokushima2008,Fransson2016}. There is a threshold for emission at $\sim$534 eV excitation energy and a pre-edge peak at 535 eV. %A partial fluorescence yield spectrum could be obtained by projecting the intensity on the vertical axis; however, due to the geometry of the experiment and jet thickness, the result does not directly represent the absorption cross section, but emphasizes weaker resonant features \cite{Eisebitt1993}. %In our geometry, the fluorescence yield spectrum is heavily saturated \cite{Eisebitt1993}, $\it{i.e.}$ it does not directly represent the absorption cross section, but put relative emphasis on weaker resonant features. 
After ionization, Fig. \ref{fig:rixsmap}(b), a new resonant feature appears at 526 eV excitation energy that is identified as $1s_O\rightarrow1\pi$ transition of OH$(aq)$: its position is near that of gas-phase OH \cite{Stranges-2002-JCP}
%In addition, the X-ray emission spectra excited at higher energies are modified compared to spectra excited before the pump pulse. 
%We now analyze the RIXS emission spectra as a function of the incident energy.  Prior to the ionization pulse, these features develop as a function of excitation energy in a way that is typical of liquid water. For high excitation energy (above 535 eV) the main emission feature is the water XES doublet in the 526-527-eV region, which is barely resolved here. %and smeared-out intensity towards lower energies as shown in the top panel of Fig. \ref{fig:XES-combo} (a). %For excitation closer to threshold the spectrum is significantly different with a narrow main feature. 
and its kinetics are consistent with proton transfer \cite{Loh2020Science}.  The position of the quasi-elastic RIXS line of OH$(aq)$ coincides with the lower-energy component of the water XES doublet (see dashed line in Fig. \ref{fig:XES-combo}), providing a check on the absolute energy and consistent with the interpretation \cite{Fuchs2008} of this peak as due to ultrafast dissociation.

%%%%%%------------------------------------Emission from OH-radical

Table \ref{tbl:TheoExp} summarises the main RIXS features of  OH$(aq)$ and OH$^-(aq)$, comparing experimental and theoretical values. 
%The emission from the 526-eV resonance in Fig. \ref{fig:XES-combo} is assigned to transitions in OH($aq$). %The OH radical has $1s_O$, $2\sigma(2s)$ and the $2\sigma(2p_z)$ orbitals filled and one vacancy in the $1\pi(p_x/p_y)$ orbital (Fig. \ref{fig:SpectraAndMOs}, bottom panel). The ground-state configuration is thus $1s_O^2 1\sigma^2 2\sigma^2 1\pi^3$, and resonant x-ray transition occurs when an electron from the core level fills the $\pi$  vacancy, yielding the $1s_O^1 1\sigma^2 2\sigma^2 1\pi^4$ state. 
The RIXS spectrum has a peak with 0.7 eV FWHM that assigned to quasi-elastic $GS\rightarrow 1s_O^{-1}\pi^{+1}\rightarrow GS$ scattering to the electronic ground state. We also observe a 1.5 eV wide structure beginning at 3.6 eV energy loss that corresponds to scattering to the first electronically excited state: $GS\rightarrow 1 s_O^{-1}\pi^{+1}\rightarrow 2\sigma^{-1}\pi^{+1}$.  
The calculations reproduce the gap ($\Delta E$) between the quasi-elastic and energy-loss peak well; however, the absolute position of the $1s_O\rightarrow 1\pi$ transition is 1 eV off. EOM-EE-CCSD excitation spectra for core-level transitions often exhibit systematic shifts of 0.5-1.5 eV, attributed to insufficient treatment of 
electron correlation \cite{Vidal:CVSEOM:18}, while the relative positions of the peaks are reproduced with higher accuracy. 
The intensity ratio of the two peaks stems from their $\pi$ and $\sigma$ character and is reproduced qualitatively by our calculations. 

\tabcolsep 6.0pt
\begin{table}[h]
  \centering
  \caption{Positions and relative intensities of the RIXS peaks for OH$(aq)$ and OH$^-(aq)$ as defined in the insets of Fig. \ref{fig:XES-combo}.  
    \protect\label{tbl:TheoExp}}
  \begin{tabular}{llcccc}
    \hline
  Species       & Source         &   $E$, eV   &  $\Delta E$, eV  &  Ratio \\
    \hline 
    OH($aq$)   & this work$^a$            & 526.0   &  -3.8              & 2.1 \\
    OH($aq$)   & this work$^b$     & 525.0  &  -4.0            & 2.04 \\         
    \hline
    OH$^-$($aq$) & Ref. \citenum{Fuchs2008}. & 526.5    &  -3.8     & 4.3 \\
    OH$^-$($aq$) & this work$^b$ &   526.2    &         -3.8    & 5.20  \\
    \hline
  \end{tabular}

$^a$ Experiment. $^b$ Theory. 
\end{table}

Both quasi-elastic and energy-loss peaks are broadened due to the interaction with polar solvent and to vibrational structure. As shown in Table \ref{tbl:OHgas}, the dipole moment in electronically excited OH is 14\% larger than in the ground state, suggesting larger inhomogeneous broadening for the energy-loss peak; this is confirmed by our QM/MM calculations where the effect of the solvent is treated explicitly. The analysis of structural differences between the ground X($^2\Pi_1$), 
valence excited A($^2\Sigma^+$), and core-excited states suggests longer vibrational progression for the energy-loss peak, which is confirmed by the computed FCFs (see SI). 
This trend can be rationalized by the shapes of molecular orbitals: the bonding character of $2\sigma$ orbital involved in the $GS\rightarrow 1s_O^{-1}\pi^{+1}\rightarrow 2\sigma^{-1}\pi^{+1}$ transition renders it more sensitive to vibrational excitation. %Thus, the calculations reproduce the observed larger broadening of the energy-loss peak relative to the elastic peak, although the computed widths are smaller than the experimental ones. 

The middle and bottom panels of Fig. \ref{fig:XES-combo} compare
the RIXS of OH$(aq)$ to that of OH$^-(aq)$ \cite{Fuchs2008}. The two species show very similar emission spectra, as expected from the similarity of the intermediate RIXS state. There is a significant difference in the intensities of the quasi-elastic $\pi$ peak and the energy-loss ($\sigma$) feature. %We note that experimentally measured intensity differences between $\pi$ and $\sigma$ transitions can be caused by anisotropy in the emitted radiation associated with different experimental geometries \cite{southworth_anisotropy_1991}. 
The observed $\pi/\sigma$ ratios for OH$(aq)$ and OH$^-(aq)$ are 2.1:1 and 4.3:1, compared to the calculated 2.04:1 and 5.20:1.
In OH$^-$ calculations, we assumed resonant excitation to the lowest XAS peak of solvated OH$^-$, which roughly corresponds to the transition to a diffuse $\sigma$-type orbital. The observed $\pi/\sigma$ RIXS ratio from OH$^-(aq)$  depends strongly on the nature of the intermediate state and, therefore, would be very sensitive to the excitation frequency; thus, the discrepancy between the computed values and Ref. \cite{Fuchs2008} could be due to different excitation regime. 
%We note that the observed $\pi/\sigma$ XES ratio from the OH$^-(aq)$ anion may be perturbed due to excitation to a %$\sigma$ state, e.g. $4a_1$ orbital, at 533.5 eV as was needed to isolate XES from OH$^-(aq)$ from that of the %majority water species \cite{Fuchs2008}. {\bf This is incorrect: The calculations already take into account 
%correct state}

To rationalize the apparent absence of the CT$_{core}$ transitions in RIXS, we computed valence and core-level transitions for model OH-H$_2$O structures. %following the analysis in Ref. \cite{Chipman2008,Chipman2011}.
%We computed valence and XAS transitions in model OH-H$_2$O structures to construct three-states models  for a qualitative understanding of why the CT transitions are suppressed in RIXS (see SI Section 2.5 for details). 
For the hemibonded structure, thought to be responsible for the CT$_{val}$ spectral feature in the UV-visible spectrum \cite{Chipman2008,Chipman2011}, the oscillator strength for the local X$\rightarrow$A valence transition is 5 times smaller than that of the CT$_{val}$  transition. In contrast, the oscillator strength for the CT$_{core}$ transition is $\sim$50 times smaller than that of the LT$_{core}$ due to the poor overlap of the lone pair of water with the compact $1s_O$ orbital of OH.
%Thus, three-states model explains why the RIXS spectrum is rid of CT transitions.

%%%%%%%%%%%%%%%%%%%%%%%%%%%%%%%%%%%%%%%%%%%%%%%%%%%%%%%%%%%%%%%%%%%%%%%%%%%%%%%%%%
%\section{Conclusion and Outlook}
In summary, we have measured RIXS of the short-lived hydroxyl radical in pure liquid water. At the OH resonance of 526 eV, an energy-loss feature at 3.8 eV, corresponding to the localized X$\rightarrow$A transition of OH$(aq)$, was observed. The position of the OH resonance relative to bulk water XES provides information relevant to the long-standing debate on the structural versus dynamical interpretation of water XES. \emph{Ab initio} calculations reproduce the positions, relative intensity, and broadening of the quasi-elastic and energy-loss peaks of OH($aq$) and OH$^-$($aq$) and provide insight into the relative intensities of the local and CT RIXS transitions. Time-resolved RIXS, enabled by the availability of intense, tunable ultrafast x-ray pulses from XFELs, highlights the localized transition in this transient species, which is otherwise hidden in direct UV absorption spectra. This ability to report on intrinsic electronic structure of OH rather than on the properties of the solvent and its structure (which is revealed by the CT transitions dominating the UV spectrum) represents the key advantage of RIXS, demonstrating that it may be used to track ultrafast reactions of the chemically aggressive hydroxyl radical in aqueous and potentially more complex environments.  

%Indication of the shape of the 526eV resonance in the RIXSMAP,further measurement.
%When measuring the RIXS of the 526 eV resonance we saw indications that the feature stayed on the same emission energy as we increased the incident energy. This might be due to bad monochromator resolution but it could also indicate...

%%%%%%%%%%%%%%%%%%%%%%%%%%%%%%%%%%%%%%%%%%%%%%%%%%%%%%%%%%%%%%%%%%%%%%%%%%%%%%%%%%
\section*{Acknowledgements}
This work was supported by the U.S. Department of Energy, Office of Science, Basic Energy Science, Chemical Sciences, Geosciences and Biosciences Division that supported the Argonne group under contract number DE-AC02-06CH11357. Use of the Linac Coherent Light Source (LCLS), SLAC National Accelerator Laboratory, and, resources of the Center for Nanoscale Materials (CNM), Argonne National Laboratory, are supported by the U.S. Department of Energy (DOE), Office of Science, Office of Basic Energy Sciences (BES) under Contracts DE-AC02-76SF00515 and DE-AC02-06CH11357. L.Y. acknowledges support from Laboratory Directed Research and Development (LDRD) funding from Argonne National Laboratory for conceptual design and proposal preparation. L.K., J.-E.R. acknowledge support from the Swedish Science Council (2018-04088).  L.K. was also supported by the EuXFEL.  Z.-H.L., T.D., and M.S.B.M.Y. acknowledge support from the Singapore Ministry of Education (MOE2014-T2-2-052, RG105/17 and RG109/18).   M.S. was supported by the CNRS GotoXFEL program. C.A. and R.S. were supported by the Cluster of Excellence `Advanced Imaging of Matter' of the Deutsche Forschungsgemeinschaft (DFG) - EXC 2056 - project ID 390715994. R.S. acknowledges support by the Chemical Sciences, Geosciences, and Biosciences Division, Office of Basic Energy Sciences, Office of Science, U.S. Department of Energy, Grant No. DE-SC0019451. M.L.V. and S.C. acknowledge support from DTU Chemistry (start-up PhD grant). S. C. acknowledges support from the Independent Research Fund Denmark – DFF-RP2 grant no. 7014-00258B. At USC, this work was supported by the U.S. National Science Foundation (No. CHE-1856342 to A.I.K.). A.I.K. is also a grateful recipient of the Simons Fellowship in Theoretical Physics and Mildred Dresselhaus Award from the Hamburg Centre for Ultrafast Imaging, which supported her sabbatical stay in Germany.

%I do not understand why references are not showing up...
%\bibliographystyle{aip}
\bibliography{main}

% Hiding until here! /Ludde--------------------------------------------------------------

\end{document}

\title{RIXS Reveals Hidden Local Transitions of the Aqueous OH Radical:\\
  Supplemental Information}

\author{L. Kjellsson}
\affiliation{Department of Physics and Astronomy, Uppsala University, Box 516, S-751 20 Uppsala, Sweden}

\author{K. Nanda}
\affiliation{Department of Chemistry, University of Southern California, United States}

\author{J.-E. Rubensson}
\affiliation{Department of Physics and Astronomy, Uppsala University, Box 516, S-751 20 Uppsala, Sweden}

\author{G. Doumy}
\affiliation{Chemical Sciences and Engineering Division, Argonne National Laboratory, United States}

\author{S. H. Southworth}
\affiliation{Chemical Sciences and Engineering Division, Argonne National Laboratory, United States}

\author{P. J. Ho}
\affiliation{Chemical Sciences and Engineering Division, Argonne National Laboratory, United States}

\author{A. M. March}
\affiliation{Chemical Sciences and Engineering Division, Argonne National Laboratory, United States}

\author{A. Al Haddad}
\affiliation{Chemical Sciences and Engineering Division, Argonne National Laboratory, United States}

\author{Y. Kumagai}
\affiliation{Chemical Sciences and Engineering Division, Argonne National Laboratory, United States}

\author{M.-F. Tu}
\affiliation{Chemical Sciences and Engineering Division, Argonne National Laboratory, United States}

\author{R. Schaller}
\affiliation{Center for Nanoscale Materials, Argonne National Laboratory, United States}
\affiliation{Department of Chemistry, Northwestern University, United States}

\author{T. Debnath}
\affiliation{Division of Chemistry and Biological Chemistry, Nanyang Technological University, Singapore}

\author{M. S. Bin Mohd Yusof}
\affiliation{Division of Chemistry and Biological Chemistry, Nanyang Technological University, Singapore}

\author{C. Arnold}
\affiliation{Center for Free-Electron Laser Science, DESY, Germany}
\affiliation{Department of Physics, Universit{\"a}t Hamburg, Germany}
\affiliation{Hamburg Centre for Ultrafast Imaging, Hamburg, Germany}

\author{W. F. Schlotter}
\affiliation{LCLS, SLAC National Accelerator Laboratory, United States}

\author{S. Moeller}
\affiliation{LCLS, SLAC National Accelerator Laboratory, United States}

\author{G. Coslovich}
\affiliation{LCLS, SLAC National Accelerator Laboratory, United States}

\author{J. D. Koralek}
\affiliation{LCLS, SLAC National Accelerator Laboratory, United States}

\author{M. P. Minitti}
\affiliation{LCLS, SLAC National Accelerator Laboratory, United States}

\author{M. L. Vidal}
\affiliation{DTU Chemistry - Department of Chemistry, Technical University of Denmark, DK-2800, Kongens Lyngby, Denmark}

\author{M. Simon}
\affiliation{Sorbonne Université and CNRS, Laboratoire de Chimie Physique-Matière et Rayonnement, France}

\author{R. Santra}
\affiliation{Center for Free-Electron Laser Science, DESY, Germany}
\affiliation{Department of Physics, Universit{\"a}t Hamburg, Germany}
\affiliation{Hamburg Centre for Ultrafast Imaging, Hamburg, Germany}

\author{Z.-H. Loh}
\affiliation{Division of Chemistry and Biological Chemistry, Nanyang Technological University, Singapore}

\author{S. Coriani}
\affiliation{DTU Chemistry - Department of Chemistry, Technical University of Denmark, DK-2800, Kongens Lyngby, Denmark}

\author{A. I. Krylov}
\affiliation{Department of Chemistry, University of Southern California, United States}

\author{L. Young}
\affiliation{Chemical Sciences and Engineering Division, Argonne National Laboratory, United States}
\affiliation{Department of Physics and James Franck Institute, The University of Chicago, United States}

\maketitle
\clearpage

\tableofcontents

\clearpage
\section{Experimental details}
\vspace{-0.1in}
Details  about the  liquid jet  and calibration  of the  monochromator
energy  is   described  in   the  Supplemental  Information   of  Ref.
\citenum{Loh2020Science}.   In  this  document,  we  extend  with  the
information about the x-ray emission spectrometer (XES).

% I guess  we skip the information  about timing /Ludde A  delay stage
%was used to scan  the delay of the ionizing pump  from -0.5 ps before
%the  x-ray probe  to  +1.2  ps after  it.  The  pump-probe jitter  of
%$\pm0.25$ ps FWHM was measured  every x-ray pulse using the time-tool
%described  in \cite{timetool2004}.  The spatial  overlap between  the
%pump  and probe  in  the  interaction region  was  confirmed using  a
%YAG-crystal.

The monochromator energy dispersion  was calculated from motor encoder
positions  using  the  standard  method at  the  SXR  instrument.  The
absolute  energy was  calibrated by  comparing the  absorption in  our
2-$\mu$m  thick water  jet  with  800-nm thick  water  as measured  by Ref.
\citenum{nagasaka2010}.

X-ray spectra were measured perpendicular  to the incoming beam, using
a             varied            line             spacing-grating-based
spectrometer\cite{chuang2017modularVLSSPEC}. The spectra were captured
shot-by-shot using a CCD detector. Due to limitations in data transfer
times  the  images  were  projected  into  one  dimension  before  the
readout. A dark  background was subtracted from the  detector data and
values over  a certain threshold  were counted.  The intensity  of the
pixel was discarded,  and only the position was  recorded. This method
would fail  to identify two  photons hitting  the same pixel,  this is
however unlikely  with the low  count rates in the  experiment. Cosmic
rays  with  substantially  higher   pixel  values  than  photons  were
identified and discarded.

The RIXS maps (Fig.  2 in the main text) was  created by binning x-ray
spectra  by incoming  photon  energy in  140 meV  wide  bins and  then
normalizing  by the  number  of XFEL  shots in  each  bin. The  energy
dispersion of the XES was determined  to be 0.18 eV/pixel by fitting a
first degree polynomial  to the elastic line, visible in  Fig 2 of the
main text. The absolute energy of the XES was then set by shifting the
emission scale such that the excitation  energy at 526 eV matched with
emission energy.  The CCD  detector had 2048  pixels, 204  pixels were
used for creating the RIXS maps in Fig. 2. The narrowest peak observed
during the experiment was 0.7 eV at FWHM.

XFEL shots with very low intensity, as measured shot by shot, were not
considered in the data. Considering the  RIXS maps (Fig. 2 in the main
text), a  total of  710053 XFEL shots  were measured;  after filtering
there were 365846 shots considered (127931 before the valence ionizing
pump and  237915 after the  pump). A small  region of time  (0-0.2 ps)
around the interaction region was removed. There were a total of 15054
counts  on  the XES  in  this  measurement;  5228 before  the  valence
ionizing pump and  9826 after the pump. For the  OH RIXS spectrum (Fig
1.  and Fig 3.  in the  main text), after removing low intensity shots
and selecting  a time  (0.2 ps-1.4  ps after  pump) and  energy window
(0.42 eV wide incident energy), there were 15194 shots with 448 counts
on the XES.

\clearpage
\section{Theoretical methods and computational details}

To gain insight into the nature  of main spectral features, we carried
out   electronic  structure   calculations  using   equation-of-motion
coupled-cluster       singles       and       doubles       (EOM-CCSD)
theory\cite{Krylov:EOMRev:07}  augmented  by  core-valence  separation
(CVS)  \cite{Cederbaum:CVSorig:80}  to  enable  access  to  core-level
states.   Specifically,   fc-cvs-EOM-EE-CCSD  and   fc-cvs-EOM-IP-CCSD
methods\cite{Vidal:CVSEOM:18,VidalDyson:2020,VidalDyson:2020-corr} were  used for computing
XAS and XES, respectively. For the  sake of brevity, below we skip the
`fc' prefix (denoting that the core electrons are frozen in the CCSD amplitudes) and refer to
these  methods as  cvs-EOM-EE/IP-CCSD.  Relevant  valence states  were
computed using standard EOM-CCSD (with core electrons frozen).

RIXS   spectra    were   computed    using   a    recently   developed
approach\cite{Nanda:RIXS:19} for computing  EOM-CC response properties
in  the  X-Ray  domain.   The  approach  is  based  on  recasting  the
two-photon  sum-over-states expressions  for  RIXS transition  moments
into  compact  closed-form  expressions   using  the  damped  response
formalism\cite{Nanda:2PA:14,Rehn:RIXS:ADC:2017,Faber:RIXS:2019}.   The
response equations are  then solved invoking CVS,  which decouples the
response states  from the  autoionizing continuum. The  combination of
damped response theory with CVS results in a robust convergence of the
response   equations\cite{Nanda:RIXS:19,Coriani:RIXS-CVS:2019}.    The
essential features of  the theory are summarized in  the next section;
full details can be found in Ref. \citenum{Nanda:RIXS:19}.

\subsection{RIXS theory}
The  RIXS  cross  section  ($\sigma^{RIXS}$)  is  a  function  of  the
scattering angle $\theta$\textemdash defined  as the angle between the
polarization vector of the incoming  photon and the propagation vector
of the outgoing photon\textemdash and  is given in terms of components
of the  RIXS transition strength  tensor ($S_{gf}$) between  the initial
($g$) and final ($f$) states :
\begin{eqnarray}
  \begin{split}
  \sigma_{gf}^{RIXS}(\theta) &= \frac{1}{15}\frac{\omega_{o}}{\omega_{i}}   \sum_{xy}  %\nonumber 
    \left[\left(2-\frac{1}{2}\sin^2\theta\right) S^{xy,xy}_{gf}  
     + \left(\frac{3}{4}\sin^2\theta - \frac{1}{2}\right) \left(S^{xy,yx}_{gf} + S^{xx,yy}_{gf}\right)\right],
  \end{split}
\end{eqnarray}
where $\omega_{o}$ and $\omega_{i}$ are frequencies of the emitted and
incident  photons,  and  the  indices $x$  and  $y$  denote  Cartesian
components\cite{Nanda:RIXS:19,Rehn:RIXS:ADC:2017,Coriani:RIXS-CVS:2019}.
$S^{ab,cd}_{gf}$ is given by
\begin{eqnarray}
  \label{2pa-strength}
 S^{ab,cd}_{gf} = \frac{1}{2}\left(M^{ab}_{gf} M^{cd}_{fg} + (M^{cd}_{gf})^* (M^{ab}_{fg})^*\right),
\end{eqnarray}
where                $^*$               denotes                complex
conjugation\cite{Hattig:MTM:98,Salek:CCTPA:06}.   Here, $M^{ab}_{fg}$s
are  the   RIXS  moments   given  by  the   following  sum-over-states
Kramers-Heisenberg-Dirac (KHD) expression\cite{Gelmukhanov:RIXS:99}:
\begin{eqnarray}
  \begin{split}
    &M_{fg}^{xy}\left(\omega_{i,x}, - \omega_{o,y}\right) = - \sum_{n} \left(\frac{\langle \Psi_f |\mu^{y}| \Psi_n\rangle\langle\Psi_n |\mu^{x}| \Psi_g\rangle}{\Omega_{ng} - \omega_{i,x} - i\epsilon_n}
     +~ \frac{\langle \Psi_f|\mu^{x}| \Psi_n\rangle\langle\Psi_n|\mu^{y}|\Psi_g\rangle}{\Omega_{ng} + \omega_{o,y} + i\epsilon_n}\right).
\label{rixs-khd}
  \end{split}
\end{eqnarray}
In  the KHD  expression,  $\Omega_{ng}  = E_n  -  E_g$  is the  energy
difference  between  states  $n$  and $g$  and  $i\epsilon_n$  is  the
imaginary inverse lifetime parameter for state $n$.  Within the EOM-CC
damped  response  framework, the  KHD  expression  is evaluated  using
EOM-CC energies and wave functions  and all $\epsilon_n$s are replaced
with a phenomenological damping factor $\epsilon$.

\subsection{Computational details}

In all  EOM-CCSD calculations,  we used the  6-311(2+,+)G(2df,p) basis
with uncontracted  core (oxygen), which  has been shown  to adequately
describe core-level states\cite{Nanda:RIXS:19}. The  basis is given in
Section   \ref{sec:basis}.    Below   we    denote   this   basis   as
uC-6-311(2+,+)G(2df,p).  All  (cvs)-EOM-EE-CCSD calculations  for  the
hydroxyl radical employ the UHF reference.
%Core electrons were frozen in EOM-CCSD calculations of valence states.

\begin{figure}[h!]
  \centering
  \includegraphics[width=8cm]{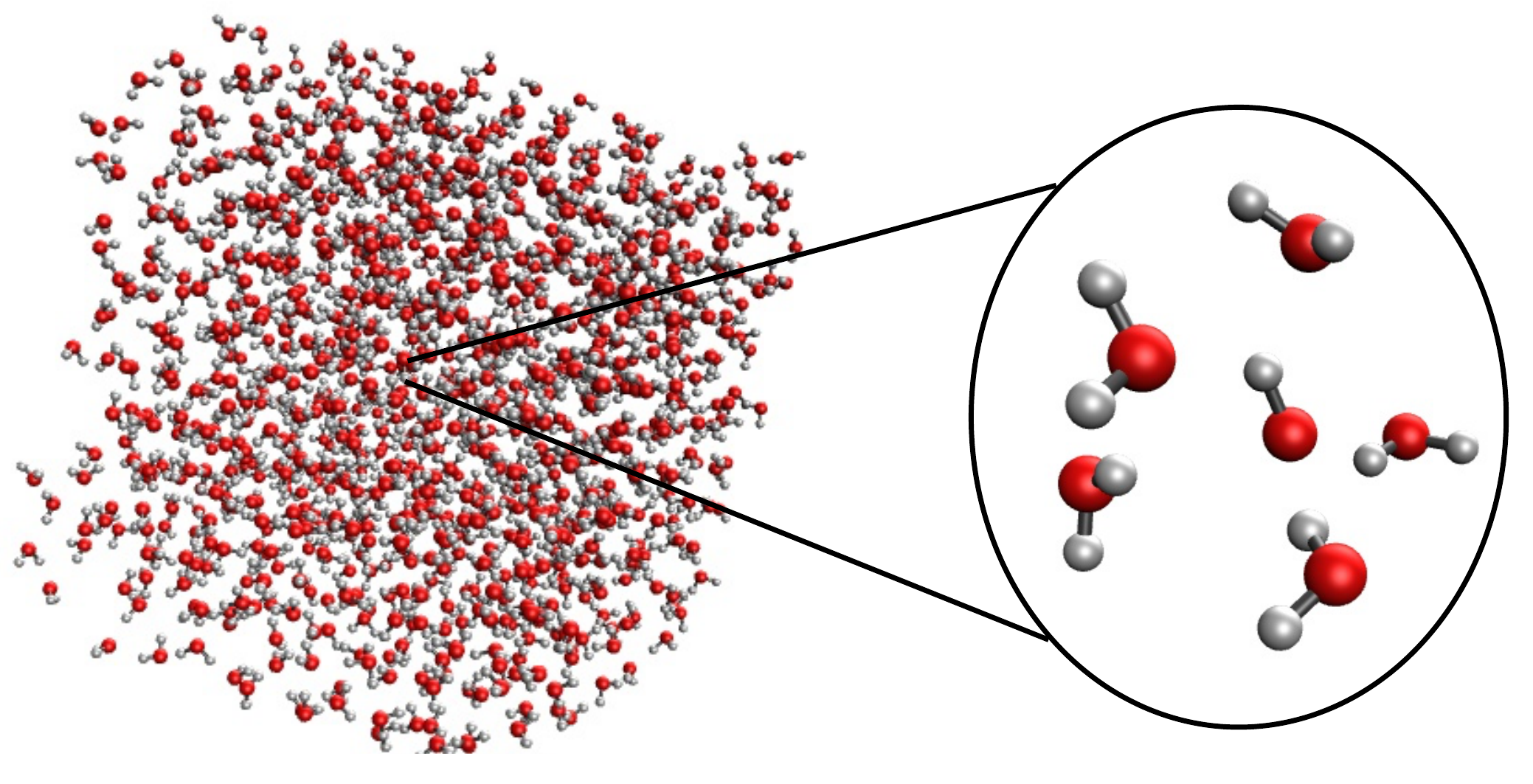}
  \caption{Snapshot from QM/MM molecular dynamics calculations.
    The QM part comprises OH (or OH$^-$) and 5 water molecules and the MM part comprises 1018 water molecules.
    In the calculations of the spectra, only OH (or OH$^-$) was included in the QM part and all 1023 waters
    were described by point charges.
     \protect\label{fig:QMMM}}
\end{figure}

We carried  out calculations  of the spectra  for isolated  species as
well as accounting for solvent  effects by using polarizable continuum
model  (PCM)  and explicit  description  of  the solvent  using  QM/MM
(quantum mechanics/molecular mechanics).  In  the XAS calculations for
isolated OH  and OH$^-$  we used  $r_{OH}$= 0.9697  \AA\ (experimental
geometry of  the hydroxyl radical\cite{Herzberg:Diatomic}).   The same
structure was used in the PCM calculations. The parameters for the PCM
calculations  were  set  up  to describe  water  ($\epsilon$=4.34  and
$\epsilon_\infty$=1.829).

To  account  for  solvent-induced spectral  shifts  and  inhomogeneous
broadening, we carried out the calculations of the spectra of solvated
species  using the  QM/MM scheme  with the  OH radical  (or hydroxide)
included in  the QM part and  with water molecules described  by point
charges   taken   from    classical   force field   ({\sl   charmm27},
Ref.  \citenum{Charmm27:FF}).    The  model  system  included   OH  or
hydroxide solvated by 1023  water molecules (see Fig. \ref{fig:QMMM}).
The  spectra  were  computed  using   100  snapshots  taken  from  the
equilibrium molecular dynamics (MD)  simulations. The computed spectra
were convoluted with  Gaussians with FWHM=0.147 eV to  account for the
life-time       broadening       of        the       $1s_O$       hole
state\cite{Stranges:XASOH:2002}.
%{\bf  Is it  ok to use  Gaussian and not Lorentzian for that?}

The starting structure for dynamics  simulations was taken from the MD
simulations of  water (courtesy of  Prof. John Herbert) and  one water
was replaced by  OH.  In the equilibrium dynamics  simulations, OH (or
OH$^-$)  and  5 water  molecules  were  included  in QM  system  (Fig.
\ref{fig:QMMM}).      The    QM     part     was    described     with
$\omega$B97X-D/6-31+G*  and the  MM  part was  described  by the  {\sl
  charmm27} force field\cite{Charmm27:FF}. In the QM/MM setup, we used
    {\sl        Janus}        interface       and        electrostatic
    embedding\cite{Thiel:Janus:2007,Shao:YingYang}.   Time  step   for
    dynamics was  42 a.u. (0.0242 fs).   Following brief equilibration
    run  of the  initial  structure,  MD was  executed  using the  NVT
    ensemble  with  T=298  K.  The trajectories  were  propagated  for
    approximately 2 ps.
%1.6 for OH and 2.4 for OHm
100 snapshots were collected after first 100 fs,
when the temperature  became stabilized. 
In the course of the simulations, the model droplet remained stable. 
%We monitored the location of QM water and size of the droplet. 
The exact setup of the dynamics simulations is illustrated by the input given at the end of this document.

To quantify the differences between valence and core-level transitions
of  local and  charge-transfer  character, we  carried out  additional
simulations on model OH-H$_2$O  structures constructed following Refs.
\citenum{Chipman:2008}    and   \citenum{Chipman:OHWater:2011};    the
respective    Cartesian    geometries    are    given    in    Section
\ref{sec:oh-h2o-structures}.

To analyze  the character of  electronic transitions, we  used natural
transition  orbitals (NTOs),  which allow  one to  describe electronic
transitions between many-body states in terms of the minimal number of
hole-electron
excitations\cite{Luzanov:TDM-1:76,HeadGordon:att_det:95,Martin:NTO:03,Luzanov:DMRev:12,Dreuw:ESSAImpl:14,Dreuw:ESSAImpl-2:14,Krylov:Libwfa:18}. We
used Gabedit\cite{Gabedit:2011} to visualize the NTOs.

To  evaluate  the  effect   of  vibrational  broadening,  we  computed
Franck-Condon  factors  (FCFs)  for the  relevant  transitions  within
double-harmonic    approximation    using   the    {\sl    ezSpectrum}
software\cite{ezspectrum}.   Frequencies and  structures for  relevant
electronic  states  of OH  (ground  state,  valence and  core  excited
states)  were computed  with (cvs)-EOM-IP-CCSD/uC-6-311(2+,+)G(2df,p).
All calculations  were carried out  using the {\sl  Q-Chem} electronic
structure     package\cite{qchem_2014,qchem_feature}.

\clearpage
\section{Results}

\subsection{Structural parameters of the OH radical}
Table  \ref{tbl:OHgas}  compares  the  theoretical  estimates  of  key
structural parameters ($r_e$, $\omega_e$, $T_e$, $\mu$) of OH with the
experimental  values\cite{Stranges:XASOH:2002}; a  shorter version  of
this table  is given as Table  1 in the main  manuscript. The computed
values  are in  good agreement  with  the experimental  ones. The  two
different    variants   of    EOM-CCSD\textemdash   EOM-EE-CCSD    and
EOM-IP-CCSD\textemdash are in agreement with each other.  We note that
the EOM-EE-CCSD values  are closer to the experimental  ones, which we
attribute to its more flexible ansatz.

\tabcolsep 6.0pt
\begin{table}[h!]
  \caption{Key structural parameters of isolated OH radical.
        \protect\label{tbl:OHgas}}
  \begin{tabular}{llccccl}
\hline    
    State    &   Character          &   $T_e$ (eV)     & $r_e$ (\AA) & $\omega_e$ (eV) & $\mu$ (a.u.) &  \\
\hline 
X($^2\Pi_1$) &  $(1s_O)^2 1\sigma^2 2\sigma^2 1\pi^3$            &  0.000           & 0.972/0.955       & 0.468/0.485   & 0.706/0.701  & $^a$ \\
             &             &  0.000           & 0.9697      & 0.463   &        & $^b$ \\ 
\hline
core $^2\Sigma^+$ & $1s_O^{-1} 1\sigma^2 2\sigma^2  \pi^4$ & 527.573/525.145  & 0.916/0.923   & 0.543/0.522   & 0.747/0.814 & $^a$ \\  
                  &                  & 525.8  & 0.915   & 0.533   &       & $^c$ \\ 
\hline
A($^2\Sigma^+$) &   $(1s_O)^2 1\sigma^2 2\sigma^1 1\pi^4$   &    4.122/4.128            & 1.014/0.976  & 0.398/0.397 & 0.808/0.801 & $^a$ \\  

                &             &   4.052             & 1.012  & 0.394   &     & $^b$ \\
\hline
  \end{tabular}

  $^a$ Theory, this work. $T_e$ and $\mu$: (cvs)-EOM-IP-CCSD/(cvs)-EOM-EE-CCSD, computed at the experimental geometries;
  $r_e$: (cvs)-EOM-IP-CCSD/B3LYP; $\omega_e$: (cvs)-EOM-IP-CCSD/B3LYP. Basis set:  uC-6-311(2+,+)G(2df,p).
$^b$ Expt., from Ref. \citenum{Herzberg:Diatomic}.
$^c$ Expt., from Ref. \citenum{Stranges:XASOH:2002}.
\end{table}

The differences  in bond lengths  and frequencies are  consistent with
the molecular orbital picture of the  electronic states (see Fig. 1 of
the main  manuscript).  We note that  the valence excited state  has a
considerably larger  dipole moment  than the ground  state ($\sim$14\%
increase),  suggesting  a  larger   inhomogeneous  broadening  of  the
energy-loss peak in the RIXS spectrum\cite{Sun:AcetoneRIXS:2011}.

\clearpage
\subsection{Calculations of XAS spectra}
\begin{figure}[t]
  \vspace{-0.in}
  \centering
  \includegraphics[width=7.5cm]{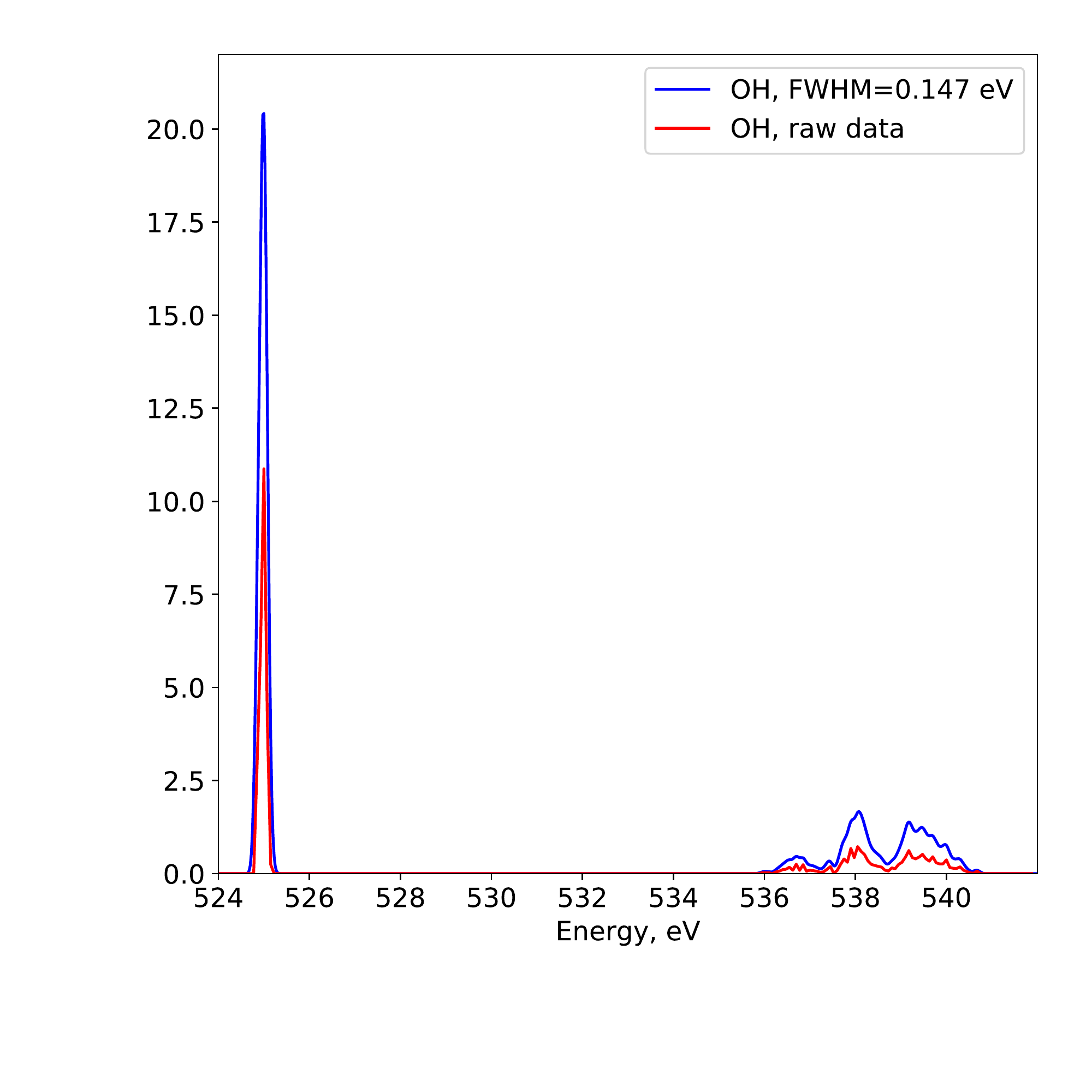}
  \includegraphics[width=7.5cm]{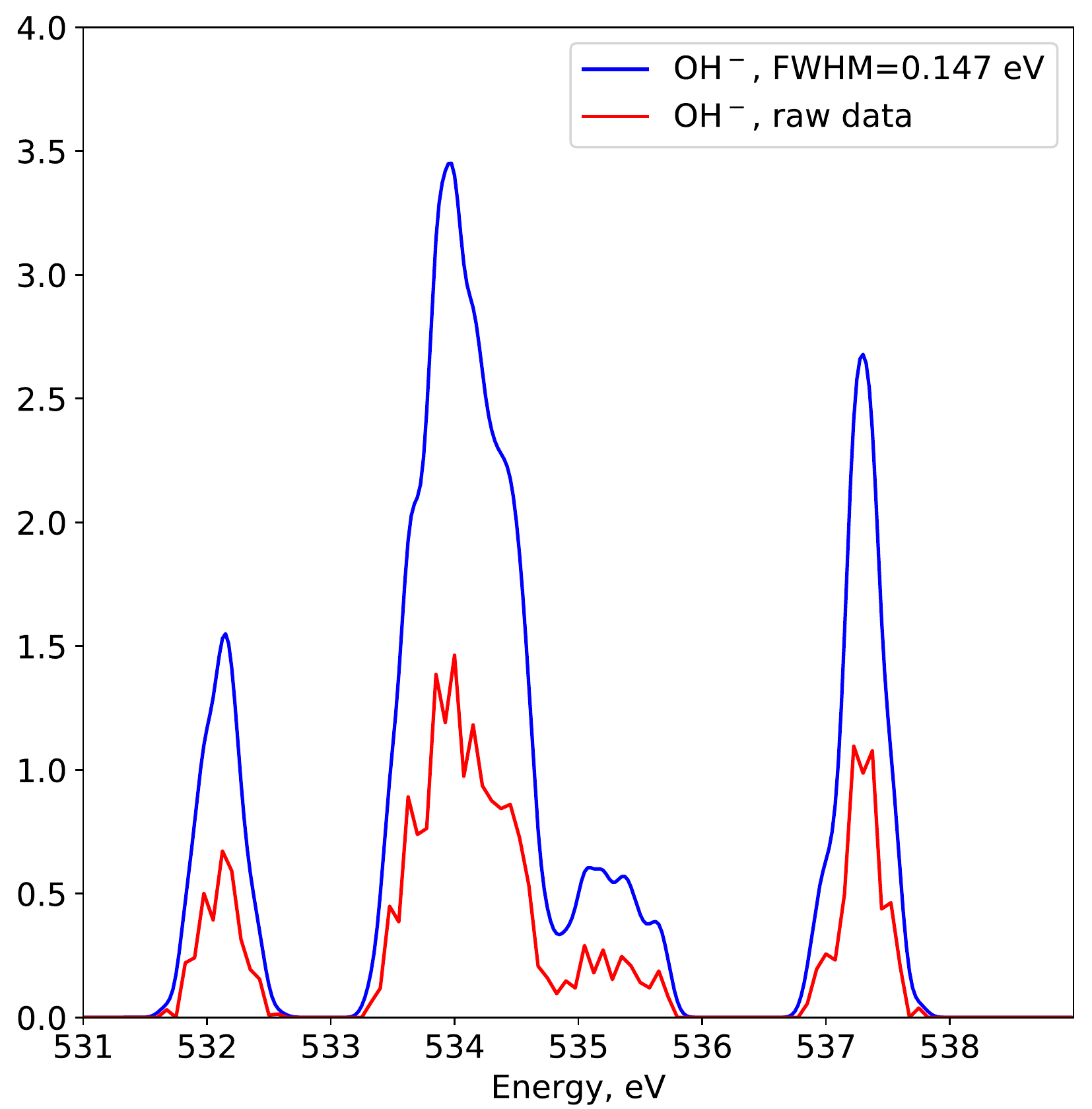}
  \vspace{-0.15in}
  \caption{Computed XAS spectrum of solvated OH (left) and OH$^-$ (right); QM/MM with cvs-EOM-EE-CCSD/uC-6-311(2+,+)G(2df,p).
       \protect\label{fig:XAS}}
  \vspace{-0.05in}
\end{figure}

\vspace{-0.15in}
Tables  \ref{tbl:XASOH}  and  \ref{tbl:XASOHm}  show  the  transitions
giving rise to  the XAS spectra of OH (vertical excitation
energies for isolated species and with PCM solvent)
and OH$^-$ (vertical excitation energies with PCM solvent), respectively.  Symmetry labels correspond to
the C$_{2v}$ subgroup used in  the calculations ($\pi$ orbitals belong
to  $b_1$ and  $b_2$ irreps;  $\sigma$ and  1$s_O$ orbitals  belong to
$a_1$). Figure \ref{fig:XAS} shows the  XAS spectra computed using the
QM/MM snapshots. We considered six lowest transitions.

\vspace{-0.05in}
\tabcolsep 6.0pt
\begin{table}[h!]
\vspace{-0.05in}
  \caption{XAS transitions in OH; cvs-EOM-EE-CCSD/uC-6-311(2+,+)G(2df,p).
    Excitation energies in eV, oscillator strengths shown in parenthesis.
    \protect\label{tbl:XASOH}}
\begin{tabular}{lcc}    
  \hline
  Transition & OH   & OH/PCM  \\
  \hline
  $b_2$  & 525.21 (0.048) & 525.17 (0.048) \\
  $a_1$  & 536.02 (0.005) & 536.52 (0.004) \\
  $a_1$  & 537.36 (0.016) & 537.83 (0.012) \\
  $b_1$  & 539.56 (0.004) & 539.06 (0.005) \\
  $b_2$  & 539.70 (0.001) & 539.41 (0.004) \\
  $b_1$  & 540.90 (0.004) & 539.64 (0.001) \\
  \hline
\end{tabular}

The hole is in $b_2$ $\pi$-orbital.
%QM/MM maxima of first and second peak are 525 and 536 eV.
%XES core-pi transition of OH- (to be compared with b2 transition):
%527.66 eV, f=0.0048 (no pcm)
\end{table}

\vspace{-0.15in}
In  the case  of  OH, the  solvent  effects appear  to  be small;  for
example, the  inclusion of the solvent  via PCM changes the  energy of
the  lowest  transition ($1s_O\rightarrow  1\pi$)  by  0.04 eV  (Table
\ref{tbl:XASOH}).  The shifts  for  other transitions  are also  quite
small.  The explicit  inclusion of the solvent via  QM/MM is necessary
for recovering  the inhomogeneous broadening;  it is also  expected to
yield more accurate estimates of solvent-induced shifts.  In agreement
with the  PCM results,  QM/MM calculations also  show a  small solvent
effect for the lowest transition:  the computed maximum for the lowest
band is 525 eV and the peak  is very narrow ($\sim$0.3 eV).  The small
effect can  be explained by  the compact  nature of the  valence $\pi$
orbital and  by relatively  weak interactions of  the neutral  OH with
water. The effect  for higher bands is larger.  We  note that in these
classical   calculations,   vibrational   broadening  is   not   fully
recovered\textemdash the zero-point  energy of OH is 2670  K, which is
an order  of magnitude more than  the average thermal energy  of OH in
the MD  calculation.  The  calculations of vibrational  broadening are
described  in  Section 3.4. %\ref{sec:FCFs}.
As clearly  seen  from  Fig.
\ref{fig:XAS} and Table \ref{tbl:XASOH}, $1s_O\rightarrow 1\pi$  has
the  largest   intensity  and  dominates  the   spectrum.   The  large
oscillator strength  is consistent  with the $s\rightarrow  p$ orbital
character of the transition (see Fig. 1 of the main manuscript).

In  contrast to  the neutral  OH, solvent  is crucially  important for
describing excited  states in  the hydroxide anion.   Without solvent,
there are no bound transitions\textemdash all excited states are shape
resonances.  The cvs-EOM-IP-CCSD IE  corresponding to $1s_O$ in OH$^-$
is 529.31  eV, considerably  lower than  543.7 eV  in the  neutral OH.
Solvent  stabilizes   the  negative   charge  and  makes   the  lowest
transitions bound (the  IE of $1s_O$ increases to 534.5  eV in the PCM
calculations).  The  difference between QM/MM  and PCM is  also larger
than in the OH case; for example, the QM/MM maximum of the lowest peak
is  at  $\sim$532  eV,  to  be  compared with  531.3  eV  in  the  PCM
calculations.   The lowest  peak corresponds  to the  transition to  a
symmetric orbital (of $a_1$ symmetry) and appears to be well separated
from the next band derived from  the excitations to $\sigma$ and $\pi$
orbitals.
% \vspace{-0.5in}
\tabcolsep 6.0pt
\begin{table}[b]
 \vspace{-0.2in}
  \caption{XAS transitions in OH$^-$; cvs-EOM-EE-CCSD/uC-6-311(2+,+)G(2df,p). Excitation energies in eV, oscillator
    strengths shown in parenthesis.
    \protect\label{tbl:XASOHm}}
\begin{tabular}{lcc}    
  \hline
  Transition & OH$^-$/PCM  \\
  \hline
 $a_1$   & 531.31 (0.006) \\
 $a_1$   & 532.79 (0.008) \\  
 $b_1/b_2$ & 532.81 (0.009) \\
  $a_1$   & 533.96 (0.005) \\
  $a_1$   & 537.07 (0.011) \\
  \hline
\end{tabular}
% QM/MM max of the first peak $\sim$532 eV.
\end{table}

\clearpage
\subsection{Calculations of RIXS spectra}

\begin{figure}[h!]
  \centering
  \includegraphics[width=14cm]{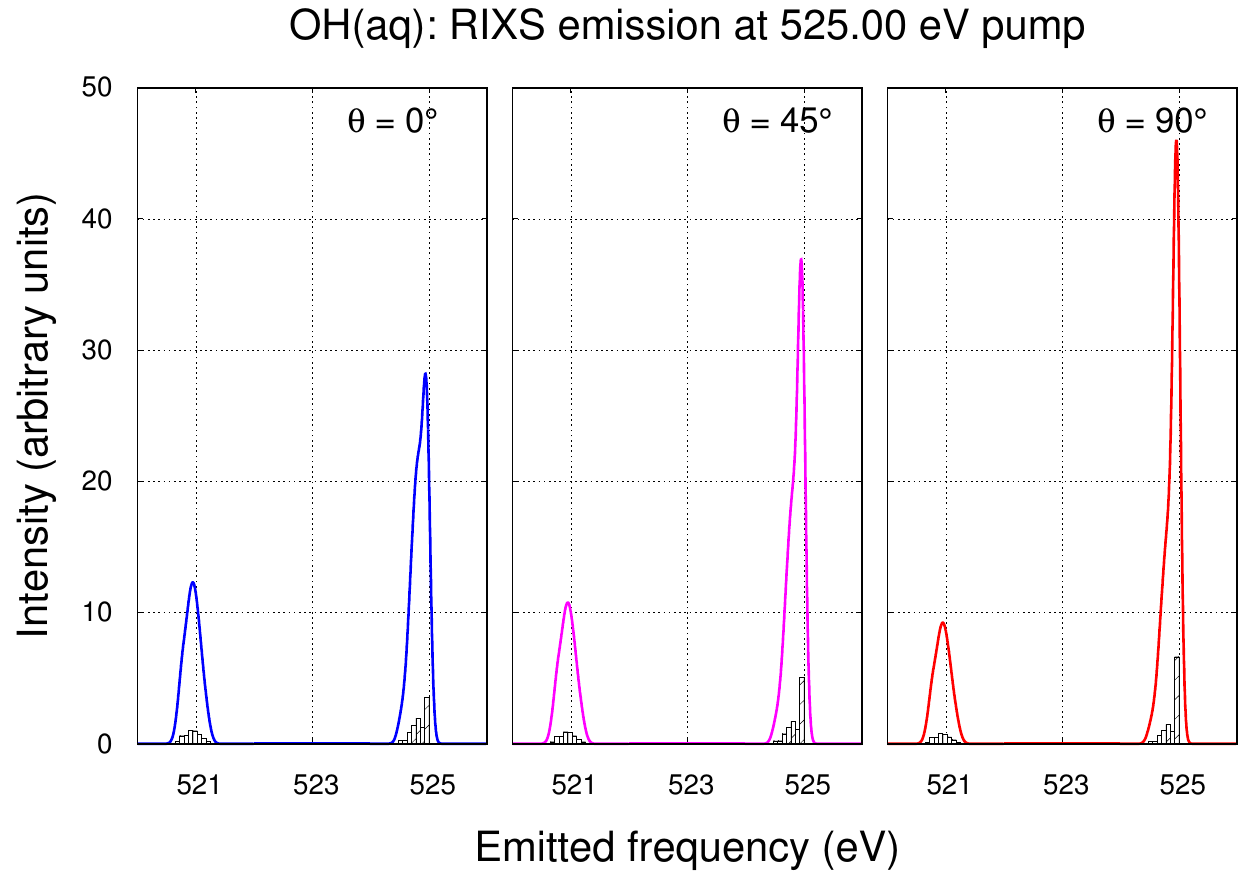}
  \caption{RIXS  spectra  of  solvated  OH computed  with  QM/MM  with
    (cvs-)EOM-EE-CCSD/uC-6-311(2+,+)G(2df,p).  The  computed transitions
    are  convoluted  with  Gaussians   with  FWHM=0.147  eV.   Pumping
    frequency is 525.00 eV.  \protect\label{fig:rixs-OH-525}}
\end{figure}

Figure \ref{fig:rixs-OH-525}  shows RIXS  spectra of  the OH($aq$) radical
computed  with three  different polarization  (the experimental  setup
corresponds to $\theta=0^\circ$).  In these calculations, lowest three
excited states of OH were included and damping factor $\epsilon$=0.005
hartree was  used. We note that higher excited states of OH cannot be
produced in the RIXS process with $1s_O\rightarrow 1\pi$ pumping transition,
because they would require  two-electron transitions. 

The computed  spectra show two peaks:  the elastic
band at  525 eV  and the  energy-loss peak  at $\sim$521  eV.  Orbital
analysis  confirms  that  the  energy-loss  peak  corresponds  to  the
relaxation of the 2$\sigma(p_z)$ electron.  The energy gap between the
two peaks is nearly identical to the XES transition computed using the
same   snapshots   and  cvs-EOM-IP-CCSD/uC-6-311(2+,+)G(2df,p).    The
relative  intensity  of the  two  peaks  depends on  polarization;  at
$\theta=0^\circ$, the intensity of the elastic peak, which corresponds
to   the  relaxation   of   doubly  degenerate   $\pi$  orbitals,   is
approximately twice higher than the intensity of the energy-loss peak.
The widths  of the  peaks are  0.32 eV (elastic)  and 0.35  eV (energy
loss).  Larger broadening  of the energy-loss peak  is consistent with
the  increased dipole  moment  of the  final  (valence excited)  state
(Table \ref{tbl:OHgas}).

\begin{figure}[h!]
  \centering
  \includegraphics[width=14cm]{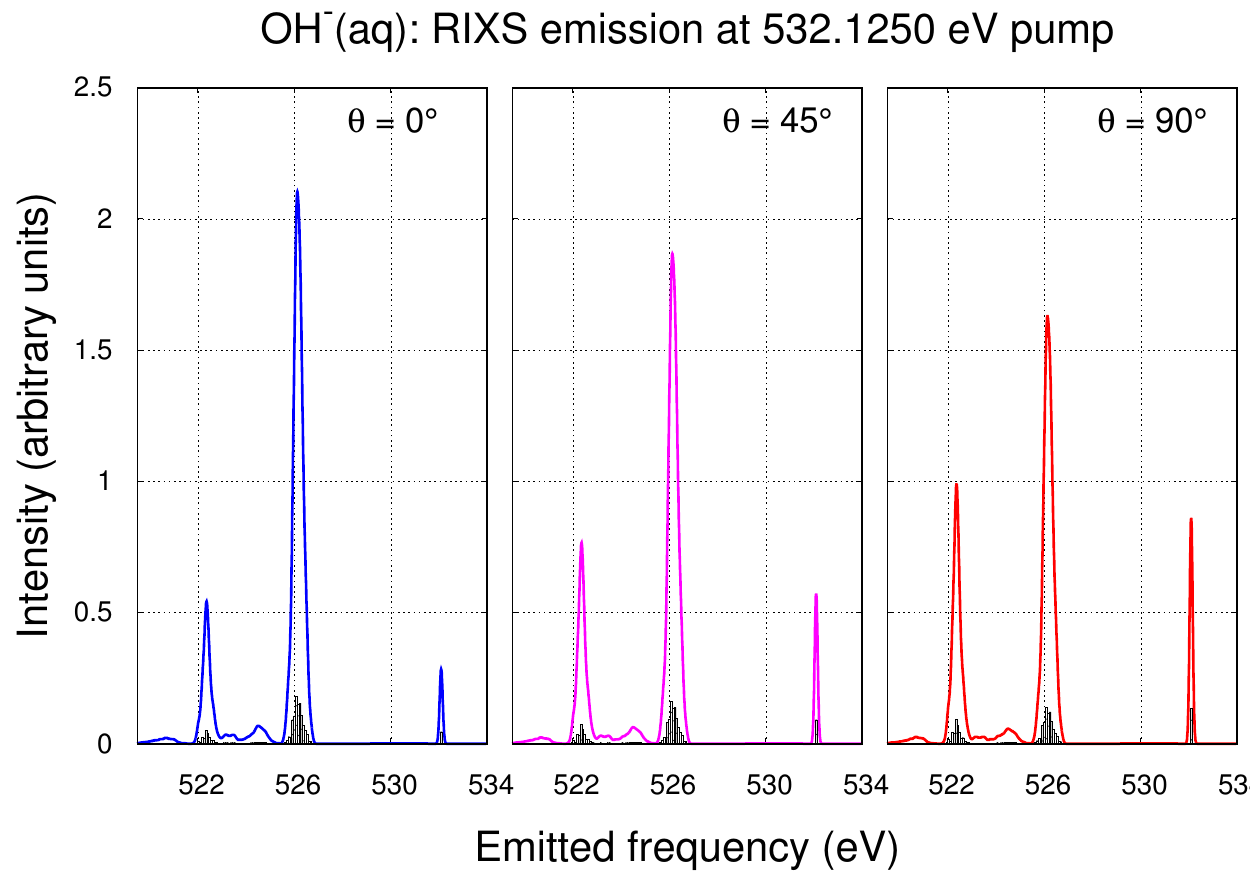}\\~\\
  \includegraphics[width=14cm]{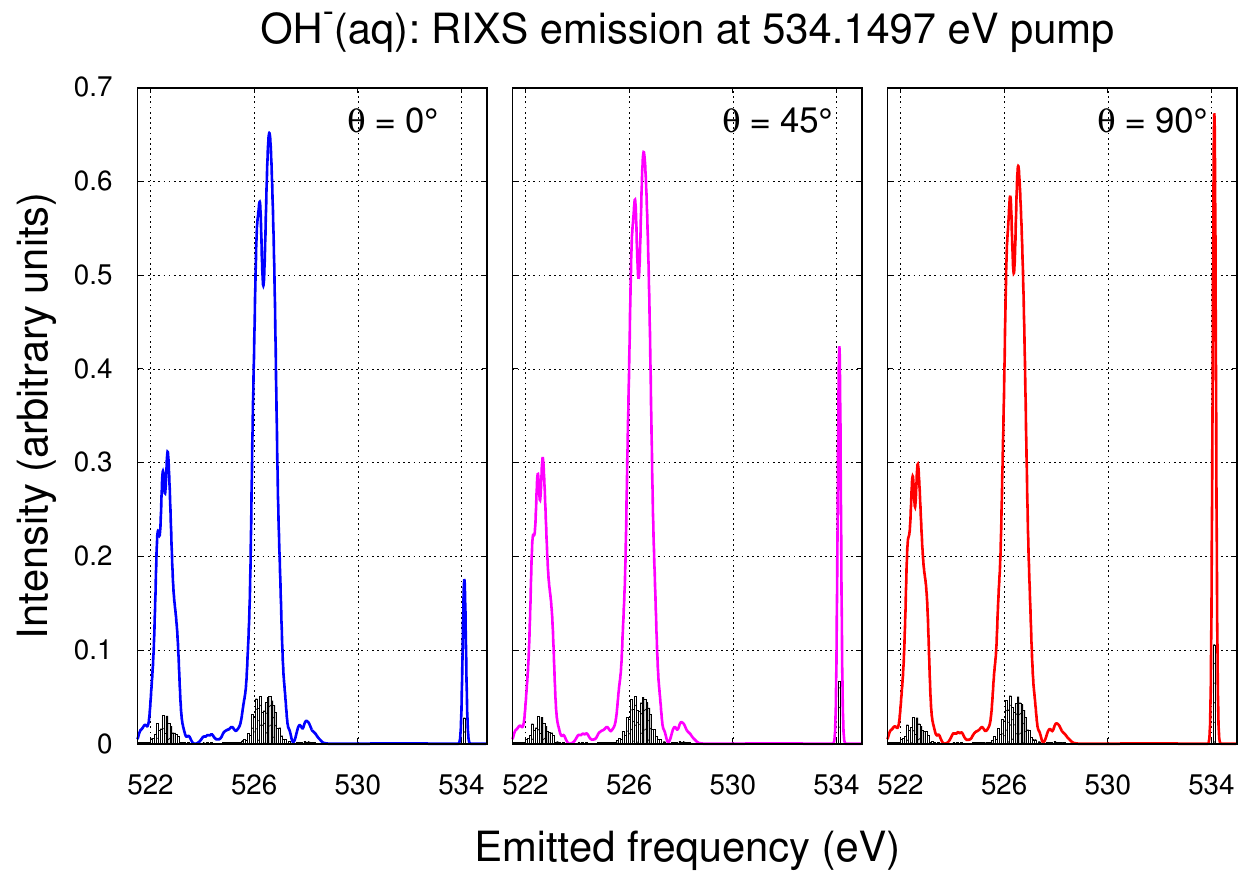}
  \caption{RIXS spectra of solvated OH$^-$ computed with QM/MM with (cvs-)EOM-EE-CCSD/uC-6-311(2+,+)G(2df,p).
    The computed transitions are convoluted with Gaussians with FWHM=0.147 eV.
    Pumping frequencies are: 532.1250 eV (top panel) and 534.1497 eV (bottom panel).
      \protect\label{fig:rixs-hydroxide}}
\end{figure}

\begin{figure}[h!]
  \centering
  \includegraphics[width=10cm]{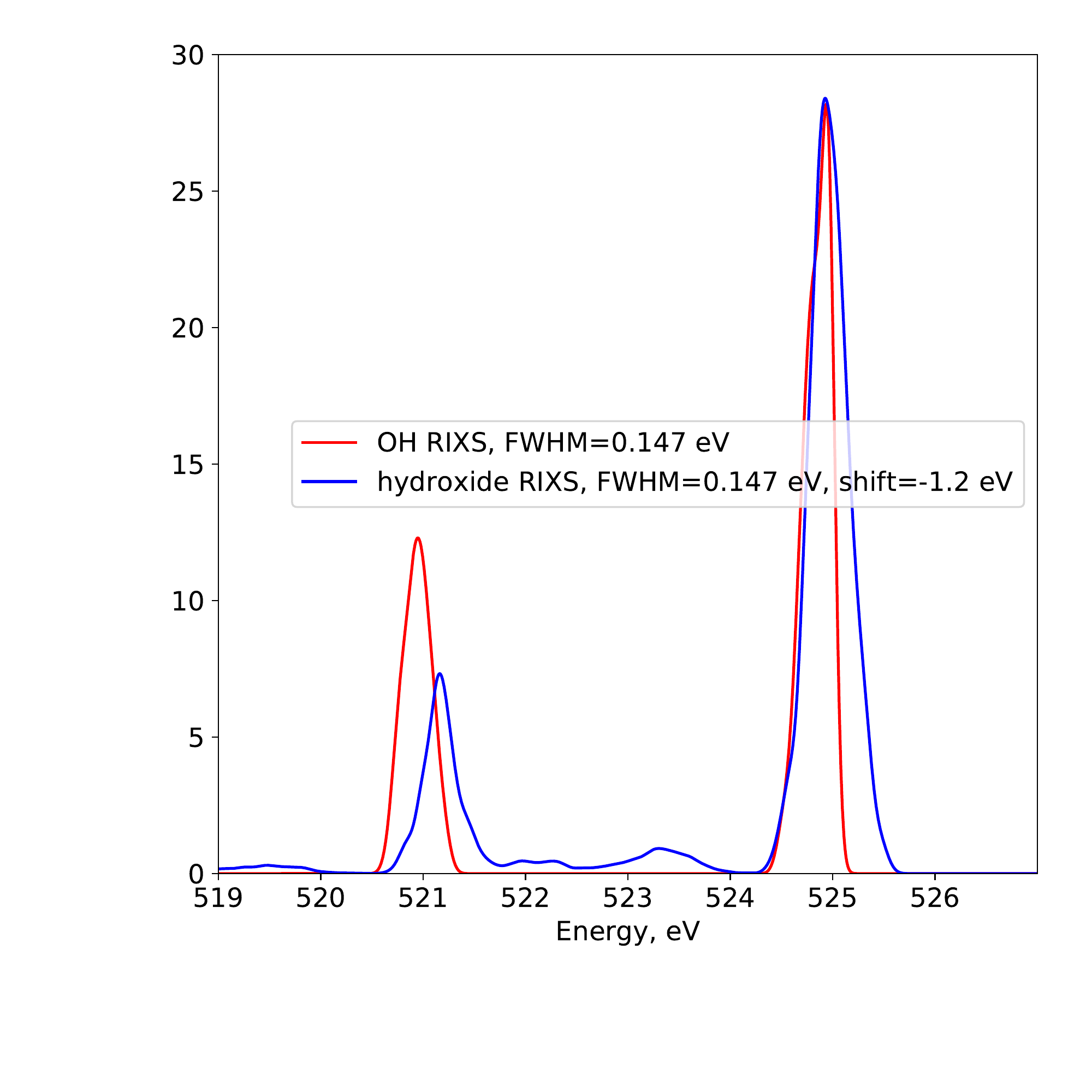}
  \caption{RIXS spectra of solvated OH and OH$^-$ computed with QM/MM with (cvs-)EOM-EE-CCSD/uC-6-311(2+,+)G(2df,p).
    The computed transitions are convoluted with Gaussians with FWHM=0.147 eV.
    $\theta=0^\circ$, pumping frequency is 525 eV for OH and 532.125 for OH$^-$. The hydroxide spectrum is blue-shifted by 1.2 eV for maximum
    alignment; intensities are arbitrary.
      \protect\label{fig:rixs-compare}}
\end{figure}

Computed RIXS emission  spectra for OH$^-{(aq)}$ are  shown in Figure
\ref{fig:rixs-hydroxide};  Fig.  \ref{fig:rixs-compare}  compares RIXS
emission  spectrum  of  OH$^-{(aq)}$  with that  of  the  OH${(aq)}$
radical.  In  these calculation,  lowest 20  excited states  of OH$^-$
were   included.   The   spectra  were   computed  with   two  pumping
frequencies: 532.1250 eV and 534.1497  eV, corresponding to the maxima
of  the two  lowest XAS  bands  (see Figure  \ref{fig:XAS}).  The  two
dominant  peaks in  all spectra  have  the same  origin as  in the  OH
spectra;   they   correspond  to   the   relaxation   of  $1\pi$   and
2$\sigma(p_z)$ electrons; their positions are also close.

The absolute  position of hydroxide's  $1\pi$ peak is  blue-shifted by
1.2 eV  relative to OH.   There are  two factors contributing  to this
value:  differences in  solvation-shell  structure around  the OH  and
OH$^-$  and  the electronic  differences  (the  presence of  spectator
electron in OH$^-$, see Fig.  3 of the main manuscript).  By computing
XES  transitions of  OH$^-$  using  the two  sets  of snapshots  (from
equilibrium   simulations  of   OH  and   OH$^-$),  we   observe  that
$\pi\rightarrow 1s_O$ transition computed  for the hydroxide snapshots
is red-shifted by 0.5 eV relative to the same transitions computed for
the hydroxyl  snapshots; thus, different solvation  is responsible for
$\sim$0.5 eV.   Thus, electronic effect  is dominant and  the computed
large  blue shift  can  be attributed  to the  presence  of the  extra
electron  in  OH$^-$.  Because  of  the  compact  nature of  the  core
orbital,  this  additional  electron  is  likely  to  destabilize  the
energies of  the valence orbitals more  that the energies of  the core
orbital,  giving  rise  to  a  blue  shift  in  $\pi\rightarrow  1s_O$
transition.  The  variations in  the peak-to-peak  energy gap  and the
widths  of  these peaks  also  reflect  the  effect of  the  spectator
electron (see Fig. 3 of  the main manuscript).  The relative intensity
of these  two peaks varies depending  on the pumping frequency  due to
different symmetries of the  intermediate state. The spectrum reported
by Fuchs {\em et~al.}\cite{Fuchs:OHmRIXS:2008} most likely corresponds
to pumping  the lowest transition  (532 eV in our  calculations).  The
relative  intensity of  the  two dominant  energy-loss  peaks in  this
spectrum is roughly  4:1 (to be compared with 2:1  in the OH spectrum)
and  the energy  gap  between them  is  slightly less  than  4 eV,  in
agreement     with    the     experimental     trend    (in     Fuchs'
spectrum\cite{Fuchs:OHmRIXS:2008}  of hydroxide,  the position  of the
energy-loss peak is 3.9 eV and the relative intensity of the two peaks
are $\sim$3:1).

%Table 2 in the main manuscript compares

%\begin{figure}[h!]
%  \centering
%  \includegraphics[width=14cm]{SI_figures/fig_ExptVTheory.png}
%  \caption{RIXS spectra of OH($aq$) and OH$^-$($aq$): Theory versus experiment.
%    The computed spectra do not include vibrational broadening.
%      \protect\label{fig:rixs-exp-theo}}
%\end{figure}

\clearpage
\subsection{Franck-Condon factors and vibrational broadening}
\protect\label{sec:FCFs}

%\begin{sidewaysfigure}[!htbp!]
%  \centering
%  \includegraphics[width=10.2cm, angle =0]{SI_figures/fig_xas-fcf-cartoon-2.pdf}
  %  \includegraphics[width=12.9cm, angle =0]{SI_figures/fig_rixs-fcf-cartoon-2.pdf}
%  \caption{Definition of the Franck-Condon factors relevant for XAS and RIXS emission spectra of OH.
%      \protect\label{fig:FCFs}}
%\end{sidewaysfigure}

\begin{figure}[!htbp!]
  \centering
\includegraphics[width=7.25cm]{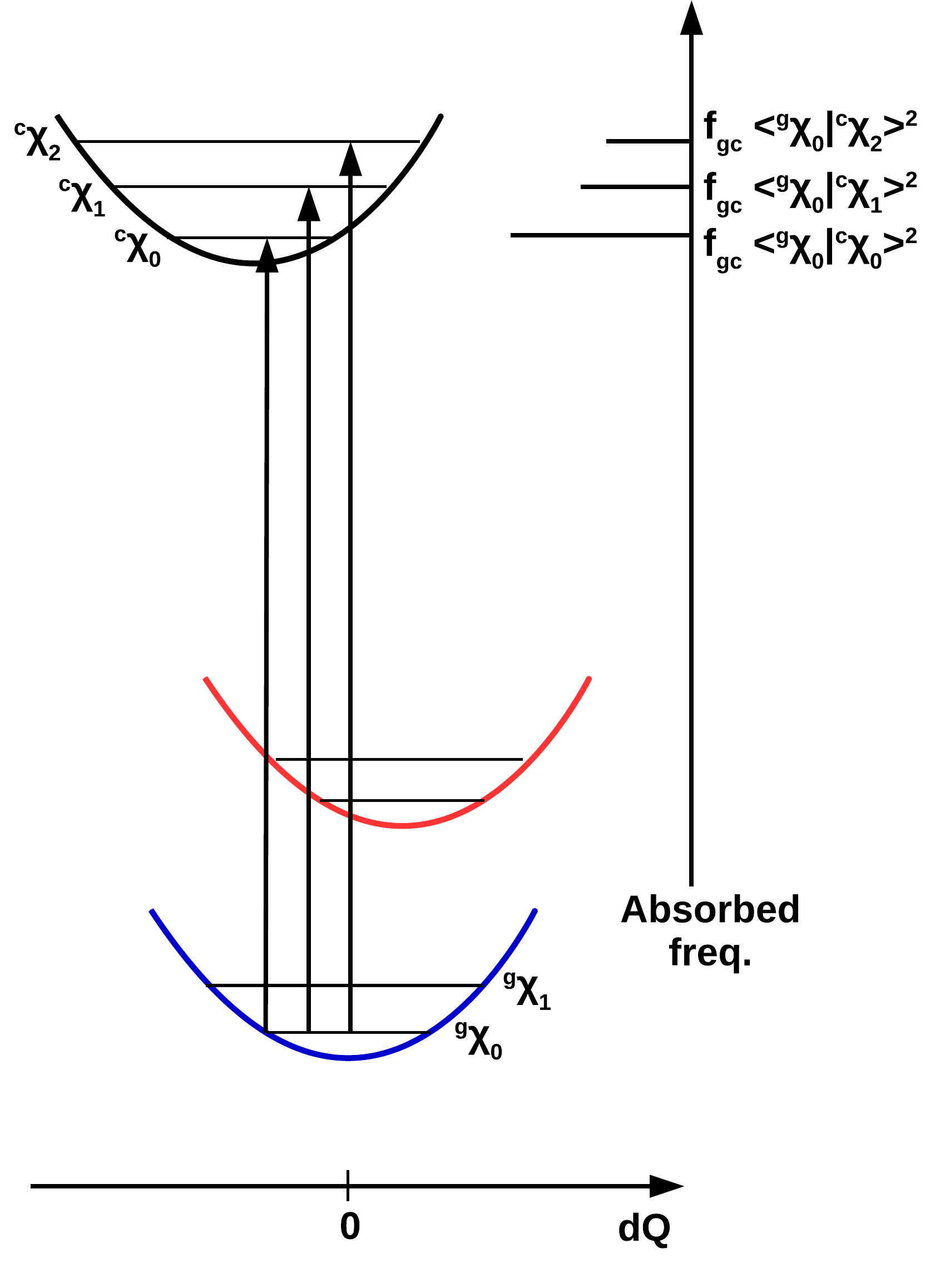}
\includegraphics[width=9cm]{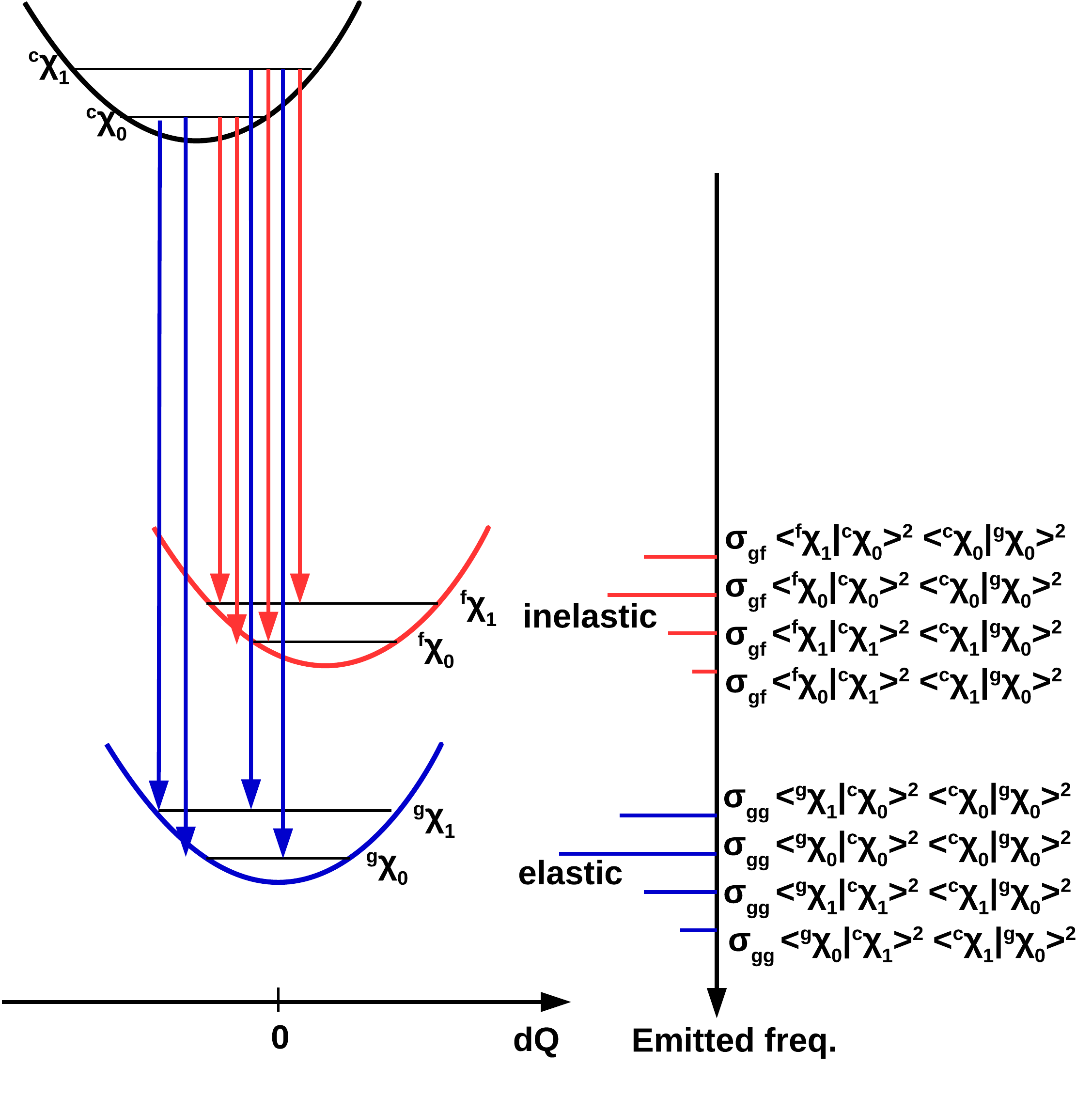}
  \caption{Definition of the Franck-Condon factors relevant for XAS (left)) and RIXS emission (right) spectra of OH.
      \protect\label{fig:FCFs}}
\end{figure}
Figure \ref{fig:FCFs} shows schematically  the potential energy curves
of the  three states relevant for  XAS and RIXS transitions.   We used
{\sl  ezSpectrum}\cite{ezspectrum}  to  compute the  FCFs  within  the
harmonic approximation  for the initial state  (X$^2\Pi_1$), the final
state     (A$^2\Sigma^+$),     and      the     core-excited     state
($(1s_O)^{-1}\ldots(1\pi)^2$)  that  is  resonant  with  the  incident
photon frequency.

\begin{figure}[h!]
  \centering
  \includegraphics[width=11cm]{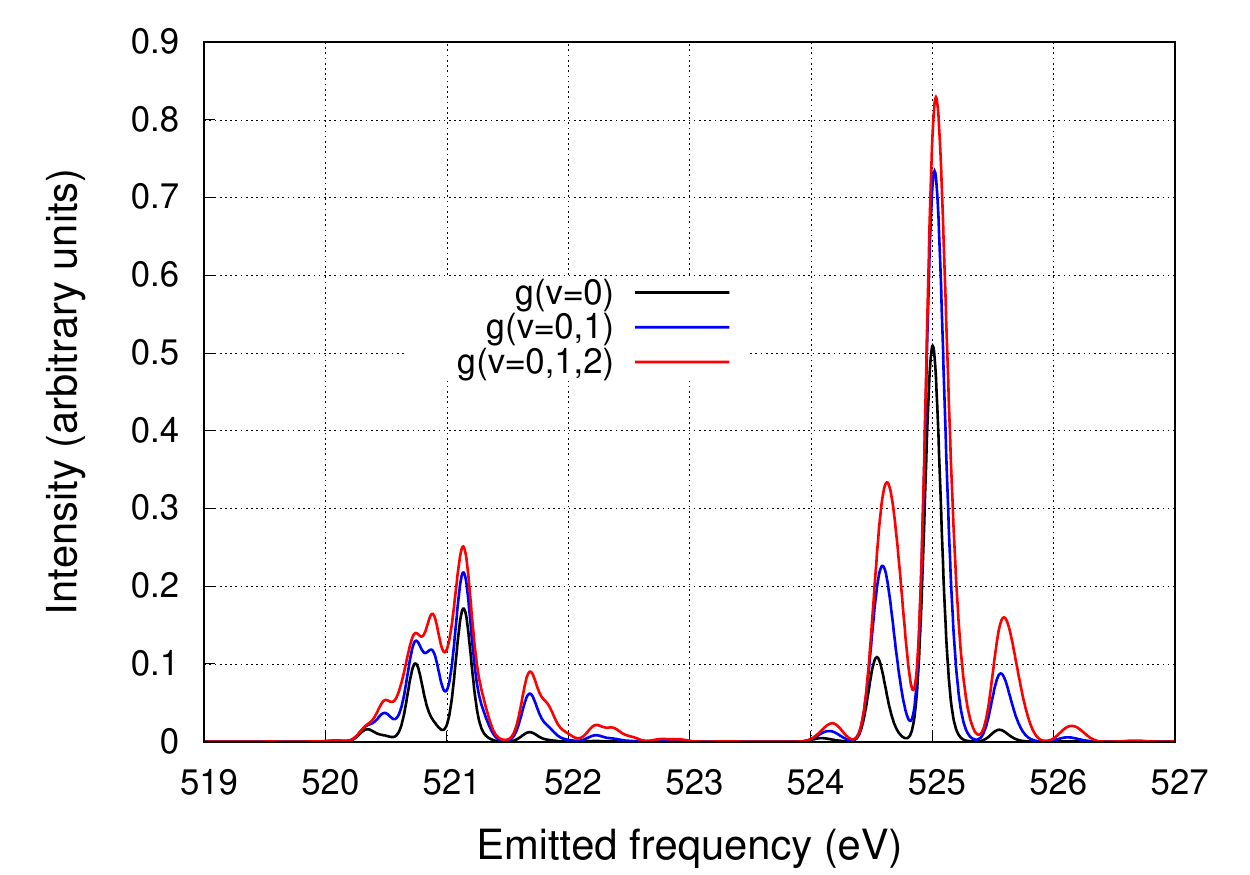}
  \includegraphics[width=11cm]{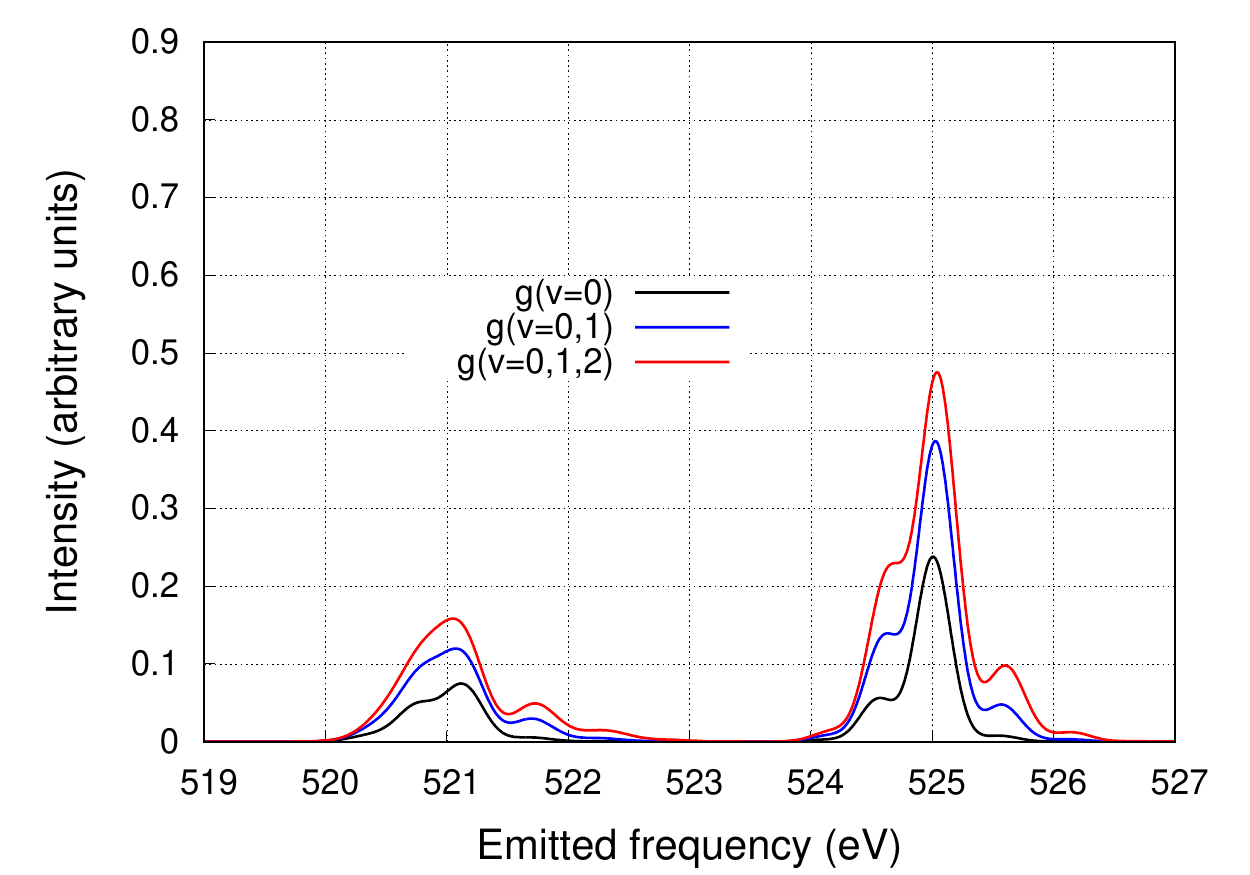}
   \caption{Computed RIXS cross sections using  harmonic FCFs smoothed with
     (top) gaussians  of FWHM=0.147 eV  and (bottom) FWHM=0.33  eV for
     the elastic  peak  and  FWHM=0.37  eV for  the inelastic  peak.
     Black curve shows the spectra computed
     assuming cold OH (only $\nu=0$ populated in the ground state).
     Blue and red curve show the spectra for hot OH with equally populated $\nu=0,1$ and
     $\nu=0,1,2$ levels, respectively.
     \protect\label{fig:FCFs-res}}
\end{figure}

%Table  \ref{tbl:OHgas} compares  the  theoretical  estimates of the
%key structural parameters ($r_e$, $\omega_e$, $T_e$, $\mu$)
%with  the experimental  values\cite{Stranges:XASOH:2002}.   The computed  values
%are in good agreement with the experimental ones.
As  one can  see  from Table  \ref{tbl:OHgas}, $1s_O\rightarrow  1\pi$
excitation results  in shorter  equilibrium distance and  an increased
vibrational frequency, as  expected from the increased  bond order due
to filling of bonding $\pi$ orbital and strong attractive potential of
the core hole.   This leads to a clear vibrational  progression in the
XAS    spectrum\cite{Stranges:XASOH:2002};     the    computed    FCFs
($\nu_0:\nu_1$  ratio =  100:16.3)  agree with  the experimental  ones
($\nu_0:\nu_1$  ratio  =  100:15.2)\cite{Stranges:XASOH:2002}.   These
FCFs  are  also responsible  for  the  vibrational broadening  of  the
elastic peak  in RIXS.  The  valence A$^2\Sigma^+$ state has  a longer
bond  length  and  softer  frequency   than  the  X$^2\Pi_1$  and  the
core-excited states,  also consistent  with their  orbital characters.
These  structural  changes give  rise  to  the  FCFs shown  in  Figure
\ref{fig:FCFs-res}. Computed FCFs are  given in Table \ref{tbl:OH:fcf}
in Section \ref{sec:fcfs}.

To compute  vibrational structure in  the RIXS spectra, we  employ the
three-states model  (in which the  sum over  all states is  reduced to
just     one    term),     similarly    to     the    treatment     in
Ref.  \citenum{Sun:AcetoneRIXS:2011}.    We  consider   the  following
scenarios:  (i) cold  OH with  only  $\nu=0$ populated  in the  ground
state; (ii) hot OH with  non-thermal populations of vibrational levels
(equal populations  of $\nu=0,1$  or $\nu=0,1,2$);  (iii) hot  OH with
thermal populations  of $\nu=0$ to $\nu=2$.   Within this three-states
model and the Condon approximation,  the RIXS scattering moments for a
specific set of  vibrational states ($p$, $q$, $r$)
for
the  initial,  intermediate,  and  final states 
are  approximated  as
follows:
\begin{eqnarray}
  \label{Eq:fcf-approx-1}
  %M_{gg}^{vib} \approx M_{gg}^{el} \sum_{pq} \langle ^g\chi_0 | ^c\chi_p \rangle \langle ^c\chi_p | ^g\chi_q \rangle; \\
  M_{gf}^{vib} \approx M_{gf}^{elec}  \langle ^g\chi_p | ^c\chi_q \rangle \langle ^c\chi_q | ^f\chi_r \rangle, 
\end{eqnarray}
where $M_{gf}^{elec}$s are given  by Eq. \eqref{rixs-khd} and computed
with  cvs-EOM-EE-CCSD.   $^n\chi_p$  represents the  vibrational  wave
function of  level $p$  for state $n$.   $\langle ^n\chi_p  | ^m\chi_q
\rangle$  is the  Franck-Condon factor  between vibrational  levels of
states $n$ and $m$.  The  RIXS scattering strengths and cross sections
for a specific set of vibrational states ($p$, $q$, $r$)
for the initial, intermediate, and final states 
are approximated using
Eq.     \eqref{2pa-strength}    for    the   electronic    part    and
Eq. \eqref{Eq:fcf-approx-1} as follows:
\begin{eqnarray}
  S_{gf}^{vib}  \approx S_{gf}^{elec} \left(\langle ^g\chi_p | ^c\chi_q \rangle \langle ^c\chi_q | ^f\chi_r \rangle \right)^2;\\
  \label{Eq:fcf-strenght-rixs}
  \sigma_{vib}^{RIXS}\left(\theta\right) \approx \sigma_{elec}^{RIXS}\left(\theta\right) \left(\langle ^g\chi_p | ^c\chi_q \rangle \langle ^c\chi_q | ^f\chi_r \rangle \right)^2.
  \label{Eq:fcf-sigma-rixs}
\end{eqnarray}
The main  difference with Ref. \citenum{Sun:AcetoneRIXS:2011}  is that
here  we  use  fixed  averaged   electronic  cross  section  for  each
vibrational transition and do not  account for small variations in the
cross sections due to variations  of the transition energies. The RIXS
emission    peak   positions    and    intensities   computed    using
Eq. \eqref{Eq:fcf-sigma-rixs} are given in Table \ref{tbl:OH:VibSplit}
in Section \ref{sec:fcfs}.

Figure \ref{fig:FCFs-res}  shows the computed RIXS  spectra. Both cold
and  hot   OH  spectra   show  vibrational   broadening.   Non-thermal
population  of  $\nu=1$  and  $\nu=2$ levels  increases  the  relative
intensity  of side  peaks, leading  to an  overall broadening  of both
elastic and inelastic bands.  When smoothened with gaussians with FWHM
extracted from  the QM/MM  simulations, one  can see  that energy-loss
peak shows  more broadening. The widths  of the peaks from  the bottom
panel of Figure \ref{fig:FCFs-res} are:
\begin{itemize}
\item For $\nu=0$: elastic peak has width of 0.345 eV and inelastic peak has width of 0.691 eV.
  This spectrum is used in the main draft for comparisons between  theory and experiment.
\item For $\nu=0,1$: elastic peak has width of 0.380 eV and inelastic has 0.726 eV.
\item For $\mu=0,1,2$: elastic peak has width of 0.458 eV and inelastic has 0.722 eV.
\end{itemize}

\clearpage
\subsection{Analysis of local and charge-transfer transitions in valence and core-level spectra}

\tabcolsep 6.0pt
%\begin{longtable}{@{}c|cccccc@{}}
%\centering
\begin{table}[b]
  \caption{Hemibonded OH + H$_2$O  complex.  Transition energies (eV),
    oscillator strengths ($f$), and  NTOs for the transitions between
    the cvs-EOM-EE-CCSD core states  and fc-EOM-EE-CCSD valence states
    with uC-6-311(2+,+)G(2df,p)  basis.  The corresponding  values for
    transitions  computed  between  cvs-EOM-IP-CCSD  core  states  and
    EOM-IP-CCSD  valence  states  are  in  parenthesis.   $\sigma_K^2$
    denotes  the  weight   of  the  corresponding  NTO   pair  in  the
    transition.          NTO        isosurface         is        0.05.
    \protect\label{SITable:CT_analysis:Hemibonded}}
  %\endfirsthead \endhead
  \begin{tabular}{@{}c|cccccc@{}}
%\scriptsize
\hline
Transition & Energy & Orb. trans. & $f$  & Hole NTO & $\sigma_K^2$ &  Particle NTO \\
\hline
\multicolumn{7}{c}{Core excitation}\\
\hline
$X\rightarrow C$  & 524.58 (527.00) & $1s_O \rightarrow 1\pi$    & 0.049 (0.054) & \centering\includegraphics[scale=0.15]{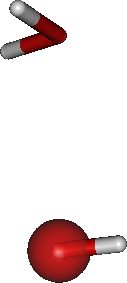} & 0.91 (1.03) & \includegraphics[scale=0.15]{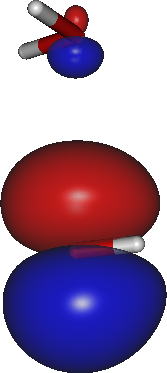}\\
\hline
\multicolumn{7}{c}{Local transitions}\\
\hline
$X\rightarrow A$  & 4.63 (4.66) & $2\sigma \rightarrow 1\pi$    & 0.009 (0.011) & \centering\includegraphics[scale=0.15]{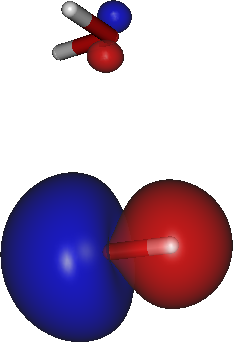} & 0.93 (0.97) & \includegraphics[scale=0.15]{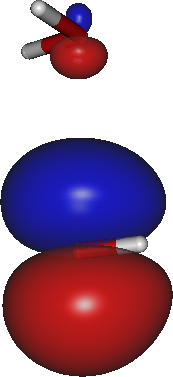}\\
\hline
$C\rightarrow A$  & 519.95 (522.34) & $2\sigma \rightarrow 1s_O$    & 0.037 (0.043) & \centering\includegraphics[scale=0.15]{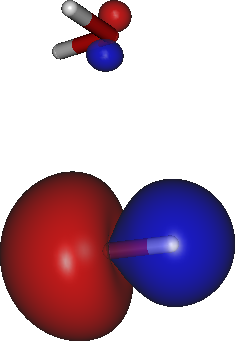} & 0.89 (1.03) & \includegraphics[scale=0.15]{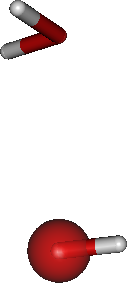}\\
\hline
\multicolumn{7}{c}{Charge-transfer transitions}\\
\hline
$X\rightarrow CT$  & 6.63 (6.08) & $lp$(H$_2$O)$\rightarrow 1\pi$    & 0.051 (0.049) & \centering\includegraphics[scale=0.15]{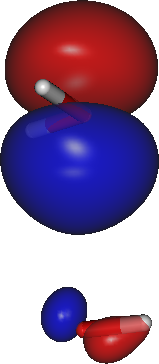} & 0.84 (0.87) & \includegraphics[scale=0.15]{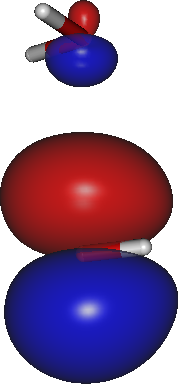}\\
\hline
$C\rightarrow CT$  & 517.95 (520.92) & $lp$(H$_2$O) $\rightarrow 1s_O$    & 0.000 (0.001) & \centering\includegraphics[scale=0.15]{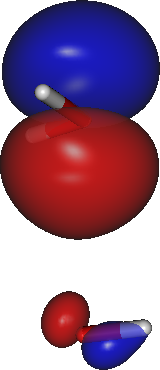} & 0.56 (0.85) & \includegraphics[scale=0.15]{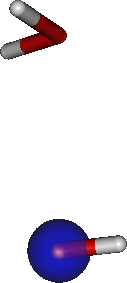}\\
\hline
\end{tabular}
\end{table}

The dominant UV-visible peaks around 5.40 eV of the aqueous OH radical
arise  due  to  charge  transfer   between  the  solvent  and  the  OH
radical\cite{Chipman:2008,Chipman:OHWater:2011}  (see  Fig. 1  of  the
main manuscript).  On the  basis of electronic structure calculations,
Ref.   \citenum{Chipman:2008} reported  that hemibonding  (or stacked)
arrangements of the OH radical with a nearby water molecule results in
charge-transfer  transitions  from  the  lone  pair  of  nearby  water
molecule  to the  singly  occupied  $1\pi$ orbital  of  OH with  large
oscillator strength,  whereas the hydrogen-bonded  acceptor structures
that also  show these charge-transfer transitions  have low oscillator
strengths. In order to understand why these charge-transfer peaks have
low  RIXS  cross   sections,  we  considered  two   model  OH+  H$_2$O
structures\textemdash one resembling a  hemibonded complex and another
an acceptor complex; the respective Cartesian coordinates are given in
Section \ref{sec:oh-h2o-structures}.

Table   \ref{SITable:CT_analysis:Hemibonded}   shows   the   energies,
oscillator strengths,  and NTOs  for one-photon  transitions involving
the  $X$,  $A$,  $CT$,  and  the  core-excited  ($C$)  states  of  the
hemibonded  structure  calculated  with  (cvs-)EOM-EE-CCSD  (and  also
(cvs-)EOM-IP-CCSD)  and using  the  uC-6-311(2+,+)G(2df,p) basis  set.
Consistent  with Ref.   \citenum{Chipman:2008}, the  $X\rightarrow CT$
transition  shows  a  larger   oscillator  strength  compared  to  the
$X\rightarrow  A$ transition.   The $\pi$-stacking  hole and  particle
NTOs for the $X\rightarrow CT$  transition show a significant overlap,
confirming its dominant  signal in the UV spectrum.   In contrast, the
$X\rightarrow  A$ transition  has a  $p_z \rightarrow  p_x$ character,
which  although  weakly  allowed,  has  negligible  transition  dipole
moment,
%along the transformation vector
giving rise to a weak signal in the UV-visible
spectrum.   The  oscillator  strength   for  the  $CT  \rightarrow  C$
transition is negligible due to poor  overlap between $1s_O$ of OH and
the lone pair  of the water molecule, whereas the  overlap between the
core  hole  and   the  particle  NTOs,  both  being   local,  for  the
$A\rightarrow C$ transition is significant.

In order  to relate properties  of these one-electron  transitions ($X
\rightarrow A$  and $X\rightarrow  CT$) with  RIXS cross  sections, we
invoke a  simple three-state  approximation of  Eq.  \eqref{rixs-khd},
similar   to    the   poor-man   RIXS   calculations    discussed   in
Ref. \citenum{Nanda:RIXS:19}.  Specifically, we  truncate the sum over
all states in Eq.  \eqref{rixs-khd}  to just one term corresponding to the core-excited
state ($C$) that is resonant  with the excitation energy\footnote{
  This   approximation  is   valid  for   two  reasons.    First,  the
  contribution  from the  off-resonant  core states  to  the sum  over
  states decays rapidly  as $\left(\Omega-\omega_i\right)^{-1}$. If the
  products  of the  transition moments  between the  off-resonant core
  states with  the initial  and final  states are  large, off-resonant
  core states cannot be omitted in the few-states model.  This, however,
  is not the case for  our model structures.  Second, the denominators
  in the sum-over-states terms  for the low-lying valence intermediate
  states  are order(s)  of magnitude  larger  than that  for the  core
  states, giving negligible contribution.  Although the sum-over-states
  terms  involving  only  the  initial   and  final  states  can  show
  significant    contributions    for    charge-transfer    two-photon
  transitions\cite{Nanda:NTO:17}, this  is not the case  for our model
  structures.  Within  the three-states  model, the numerators  in the
  RIXS scattering moment for the $X\rightarrow A$ transition are given
  as  products of  $X\rightarrow  C$ and  $C\rightarrow A$  one-photon
  moments,  which are  both  non-negligible.  On  the  other hand,  the
  numerators in the RIXS three-states  model for the $X\rightarrow CT$
  transition are  products of  $X\rightarrow C$ and  $C\rightarrow CT$
  one-photon moments, the latter  being negligible. This analysis with
  the  three-states   models  is  reflected  in   the  computed  cross
  sections.}.
%Still not sure, may be move back into text, but to the end of this section.
%I agree with the content, but it is somewhat out of place here and distracts from
%the main line
The validity of this approximation, which is justified by the resonant
nature  of RIXS  process,  is  supported by  the  RIXS cross  sections
presented  in  Table  \ref{SITable:CT_analysis:Hemibonded:RIXS}.   The
results  show that  RIXS cross  sections are  two orders  of magnitude
smaller   for  the   $X\rightarrow   CT$  transition   than  for   the
$X\rightarrow A$ transition.

\tabcolsep 6.0pt
%\begin{longtable}{@{}c|cccccc@{}}
%\centering
\begin{table}[h]
  \caption{Energy  loss  (eV)  and  RIXS cross  sections  (a.u.)   for
    hemibonded OH + H$_2$O complex computed with the (cvs-)EOM-EE-CCSD
    method  and  uC-6-311(2+,+)G(2df,p)  basis.  $\theta  =  0^\circ$.
    \protect\label{SITable:CT_analysis:Hemibonded:RIXS}}
  %\endfirsthead \endhead
  \begin{tabular}{@{}c|cccccc@{}}
%\scriptsize
\hline
Transition & Energy loss & RIXS cross section \\
\hline
Elastic & 0.00 & 0.0386 \\
$\approx$ Elastic & 0.18 & 0.0763 \\
$X\rightarrow A$ & 4.63 & 0.0585 \\
$X\rightarrow CT$  & 6.63 & 0.0006 \\
\hline
\end{tabular}
\end{table}

The  computed RIXS  cross  sections for  the hydrogen-bonded  acceptor
complex            are            given            in            Table
\ref{SITable:CT_analysis:Acceptor:RIXS}. The trend in these RIXS cross
sections,  which is  similar to  that for  the hemibonded  complex, is
consistent  with   the  corresponding  three-states  models   for  the
$X\rightarrow A$ and $X\rightarrow CT$ transitions and originates from
the negligible oscillator strength of the $C\rightarrow CT$ transition
relative   to    the   $C\rightarrow   A$   transition    (see   Table
\ref{SITable:CT_analysis:Acceptor}).  In addition,  in contrast to the
hemibonded structure, the hydrogen-bonded  acceptor complex also shows
a negligible oscillator strength for the $X\rightarrow CT$ transition.
Thus, the  contribution of  this transition  to both  the UV  and RIXS
spectra is negligible, consistent with Ref. \citenum{Chipman:2008}.

\tabcolsep 6.0pt
%\begin{longtable}{@{}c|cccccc@{}}
%\centering
\begin{table}[h]
  \caption{Energy  loss  (eV)  and  RIXS cross  sections  (a.u.)   for
    hydrogen-bonded  acceptor OH  + H$_2$O  complex computed  with the
    (cvs-)EOM-EE-CCSD method  and uC-6-311(2+,+)G(2df,p)  basis.  $\theta =
    0^\circ$.  \protect\label{SITable:CT_analysis:Acceptor:RIXS}}
  %\endfirsthead \endhead
  \begin{tabular}{@{}c|cccccc@{}}
%\scriptsize
\hline
Transition & Energy loss & RIXS cross section \\
\hline
Elastic & 0.00 & 0.0377  \\
$\approx$ Elastic & 0.14 & 0.0741  \\
$X\rightarrow  A$ & 4.15 & 0.0525 \\
$X\rightarrow CT$  & 5.33 & 0.0005 \\
\hline
\end{tabular}
\end{table}

\tabcolsep 6.0pt
%\begin{longtable}{@{}c|cccccc@{}}
%\centering
\begin{table}[h]
  \caption{Hydrogen-bonded acceptor  OH + H$_2$O  complex.  Transition
    energies  (eV),  oscillator strengths ($f$),  and  NTOs for  the
    transitions between cvs-EOM-EE-CCSD core states and fc-EOM-EE-CCSD
    valence   states    with   uC-6-311(2+,+)G(2df,p)    basis.    The
    corresponding    values   for    transitions   computed    between
    cvs-EOM-IP-CCSD core states and  EOM-IP-CCSD valence states are in
    parenthesis.    $\sigma_K^2$   represents   the  weight   of   the
    corresponding  NTO   pair  to  the  electronic   transition.   NTO
    isosurface is 0.05.  \protect\label{SITable:CT_analysis:Acceptor}}
  %\endfirsthead \endhead
  \begin{tabular}{@{}c|cccccc@{}}
%\scriptsize
\hline
Transition & Energy & Orb. trans. & $f$  & Hole NTO & $\sigma_K^2$ &  Particle NTO \\
\hline
\multicolumn{7}{c}{Core excitation}\\
\hline
$X\rightarrow C$  & 525.22 (527.59) & $1s_O \rightarrow 1\pi$    & 0.048 (0.054) & \centering\includegraphics[scale=0.15]{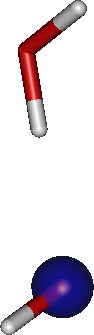} & 0.91 (1.04) & \includegraphics[scale=0.15]{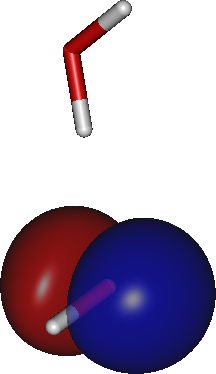}\\
\hline
\multicolumn{7}{c}{Local transitions}\\
\hline
$X\rightarrow A$  & 4.15 (4.12) & $2\sigma \rightarrow 1\pi$    & 0.002 (0.002) & \centering\includegraphics[scale=0.15]{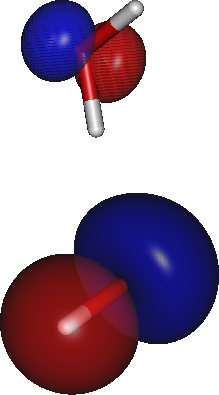} & 0.92 (0.96) & \includegraphics[scale=0.15]{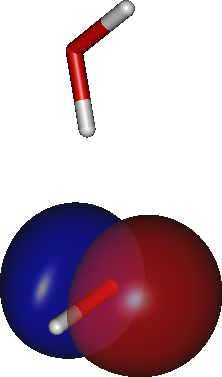}\\
\hline
$C\rightarrow A$  & 521.07 (523.47) & $2\sigma \rightarrow 1s_O$    & 0.034 (0.037) & \centering\includegraphics[scale=0.15]{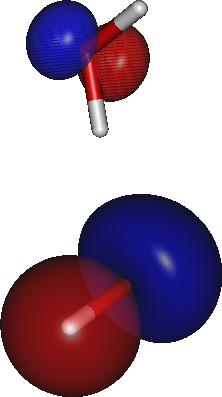} & 0.87 (1.01) & \includegraphics[scale=0.15]{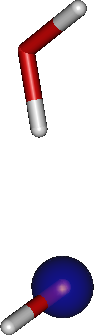}\\
\hline
\multicolumn{7}{c}{Charge-transfer transitions}\\
\hline
$X\rightarrow CT$  & 5.33 (4.58) & $lp$(H$_2$O)$\rightarrow 1\pi$    & 0.000 (0.000) & \centering\includegraphics[scale=0.15]{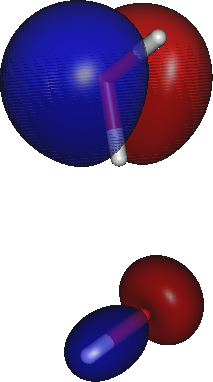} & 0.82 (0.84) & \includegraphics[scale=0.15]{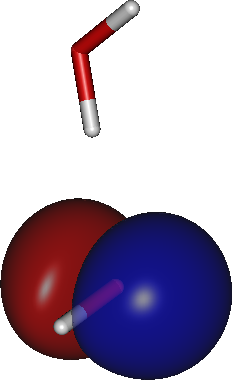}\\
\hline
$C\rightarrow CT$  & 519.88 (523.00) & $lp$(H$_2$O) $\rightarrow 1s_O$    & 0.000 (0.003) & \centering\includegraphics[scale=0.15]{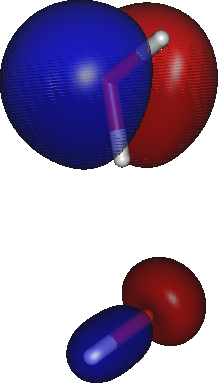} & 0.61 (0.84) & \includegraphics[scale=0.15]{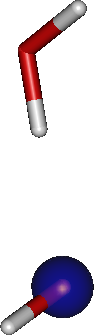}\\
\hline
\end{tabular}
\end{table}

%Keep prf, not prl for now -- to include titles 
\clearpage
%\section{References}
%\printbibliography

%\nocite{*}
\bibliographystyle{prf}

%\bibliography{abbr,general,cs,cc,dyn,krylovgroup,ab_initio,original,pyp,epif-dnabases,hf,water,na_bases,epif-gfp,chem,exp,epif,pypgfpwater,nodimer,DIP,kbravaya,ri_cholesky,tensor,web,solar,nonad,benzene,ET,sochem,so,standards,mfocc,vv,mivanov,arik,tpa,LR01OH-noDOI}

\clearpage
\renewcommand{\baselinestretch}{1.0}

\section{Input for QM/MM AIMD simulations}

\begin{verbatim}
$rem
   JOB_TYPE  aimd
   BASIS     6-31+G*
   METHOD    wB97XD
   scf_convergence  8
   scf_algorithm    gdm     does not converge with defaults
   scf_max_cycles   200
   thresh           14      use tight thresholds 
!QM/MM keywords
   qm_mm_interface janus
   user_connect    true
   force_field     charmm27 
   model_system_mult    1
   model_system_charge -1
! aimd keywords
time_step	42    in au, 1 a.u. = 0.0242 fs
aimd_steps	2000   about 2 ps
aimd_init_veloc	thermal
aimd_temp	298    
! Thermostat
aimd_thermostat         langevin
aimd_langevin_timescale 100
! From JMH input
sym_ignore      true
no_reorient     true
chelpg          true
chelpg_dx       10
chelpg_head     30
chelpg_H        50
chelpg_HA       50
mm_subtractive  true    Ewald requires mm_subtractive
ewald_on        true
$end

$forceman
ewald
alpha .274 .0649
box_length 31.3192 31.3192 31.3192
$end

$qm_atoms
1:17
$end
\end{verbatim}

\section{Input for QM/MM RIXS calculations}

\begin{verbatim}
$rem
   BASIS  =  general
   METHOD  =  eom-ccsd
   ee_states = [10]
   cvs_ee_states = [0]
   n_frozen_core = 1
   cc_fullresponse 0
   cc_eom_rixs = 5
   CC_REF_PROP 1
   CC_EOM_PROP 1
   cc_trans_prop = 1
   cc_diis_size = 15   for better convergence of response equations
   scf_algorithm = gdm   does not converge with defaults
   scf_guess =  CORE
   max_scf_cycles = 200
   thresh =  14         tight thresholds for integrals  
!QM/MM keywords
   qm_mm_interface = janus
   force_field = charmm27
   user_connect = true
   model_system_mult = 2
   model_system_charge = 0
$end

$rixs
omega_1 4234411 500 1 0
damped_epsilon 0.005
$end

$qm_atoms
1 2
$end

$molecule
0 2
O   -2.0531082762   0.70286735   -0.1205079041   101   2   0   0   0
H   -2.551746298   1.5791146684   -0.3299733637   88   0   0   0   0
O   -2.5462451975   -0.0172690537   2.6717170398   101   4   5   0   0
H   -3.0521231806   -0.8313175012   2.8204185203   88   3   0   0   0
H   -2.3438821345   0.065630876   1.7224541544   88   3   0   0   0
O   -4.0784774935   -1.0408141164   -0.9557557305   101   7   8   0   0
H   -4.0119399393   -0.5415917412   -0.1213427646   88   6   0   0   0
H   -3.5909641541   -1.8633962833   -0.7384154628   88   6   0   0   0
...
...
$end
\end{verbatim}

\section{uC-6-311(2+,+)G(2df,p) basis}
\protect\label{sec:basis}
%\vspace{-0.1in}
\begin{verbatim}
O    0
S    1    1.000000
   8.58850000E+03    1.89515000E-03 
S    1    1.000000
   1.29723000E+03    1.43859000E-02 
S    1    1.000000
   2.99296000E+02    7.07320000E-02 
S    1    1.000000
   8.73771000E+01    2.40001000E-01 
S    1    1.000000
   2.56789000E+01    5.94797000E-01 
S    1    1.000000
   3.74004000E+00    2.80802000E-01 
SP   3    1.000000
   4.21175000E+01    1.13889000E-01   3.65114000E-02 
   9.62837000E+00    9.20811000E-01   2.37153000E-01 
   2.85332000E+00   -3.27447000E-03   8.19702000E-01 
SP   1    1.000000
   9.05661000E-01    1.00000000E+00   1.00000000E+00 
SP   1    1.000000
   2.55611000E-01    1.00000000E+00   1.00000000E+00 
SP   1    1.000000
   8.45000000E-02    1.00000000E+00   1.00000000E+00 
SP   1    1.000000
   2.54518072E-02    1.00000000E+00   1.00000000E+00 
D    1    1.000000
   2.58400000E+00    1.00000000E+00 
D    1    1.000000
   6.46000000E-01    1.00000000E+00 
F    1    1.000000
   1.40000000E+00    1.00000000E+00 
****
H    0
S    3    1.000000
   3.38650000E+01    2.54938000E-02 
   5.09479000E+00    1.90373000E-01 
   1.15879000E+00    8.52161000E-01 
S    1    1.000000
   3.25840000E-01    1.00000000E+00 
S    1    1.000000
   1.02741000E-01    1.00000000E+00 
S    1    1.000000
   3.60000000E-02    1.00000000E+00 
P    1    1.000000
   7.50000000E-01    1.00000000E+00 
****
\end{verbatim}

\section{Cartesian coordinates for model OH-water structures}
\protect\label{sec:oh-h2o-structures}

\begin{verbatim}
Hemibonded structure
          O       1.4689599962    -0.0675854865     0.0543544388
          H       1.4892207572     0.5364418801    -0.3999323009
          O      -1.3070289695     0.0556311190    -0.1037339247
          H      -1.4210921138     0.4176221150     0.7621320309
          H      -1.3635768572    -0.8584290545     0.0328361571
  Nuclear Repulsion Energy =          31.83900810 hartrees       
\end{verbatim}
\begin{verbatim}
Hydrogen-bonded acceptor structure
          O       1.5259508055    -0.1019172743     0.0560387357
          H       1.8816264682     0.6459120289    -0.4385043272
          O      -1.4611163710     0.0854104501     0.0769981233
          H      -1.9178709826    -0.5428697210    -0.5149135605
          H      -0.4824309615     0.0290122853    -0.1108769843
  Nuclear Repulsion Energy =          29.49383243 hartrees
\end{verbatim}

\clearpage
\section{Franck-Condon factors}
\protect\label{sec:fcfs}

\renewcommand{\baselinestretch}{1.0}
\tabcolsep 10.0pt
\begin{longtable}{@{}l|c@{}}
  \caption{Franck-Condon factors for relevant electronic transitions in OH radical.
    \protect\label{tbl:OH:fcf}}
  \scriptsize
  \endfirsthead \endhead
%  \begin{tabular}{l|c}
\hline    
Label & FCF \\
\hline    
$\langle ^c\chi_0 | ^g\chi_0 \rangle$ & 0.913\\
$\langle ^c\chi_0 | ^g\chi_1 \rangle$ & -0.402\\
$\langle ^c\chi_0 | ^g\chi_2 \rangle$ & 0.077\\
$\langle ^c\chi_0 | ^g\chi_3 \rangle$ & 0.005\\
$\langle ^c\chi_0 | ^g\chi_4 \rangle$ & -0.006\\
\hline    
$\langle ^c\chi_1 | ^g\chi_0 \rangle$ & 0.373\\
$\langle ^c\chi_1 | ^g\chi_1 \rangle$ & 0.746\\
$\langle ^c\chi_1 | ^g\chi_2 \rangle$ & -0.535\\
$\langle ^c\chi_1 | ^g\chi_3 \rangle$ & 0.135\\
$\langle ^c\chi_1 | ^g\chi_4 \rangle$ & 0.007\\
\hline    
$\langle ^c\chi_2 | ^g\chi_0 \rangle$ & 0.156\\
$\langle ^c\chi_2 | ^g\chi_1 \rangle$ & 0.457\\
$\langle ^c\chi_2 | ^g\chi_2 \rangle$ & 0.594\\
$\langle ^c\chi_2 | ^g\chi_3 \rangle$ & 0.614\\
$\langle ^c\chi_2 | ^g\chi_4 \rangle$ & 0.192\\
\hline    
$\langle ^c\chi_3 | ^g\chi_0 \rangle$ & 0.059\\
$\langle ^c\chi_3 | ^g\chi_1 \rangle$ & 0.243\\
$\langle ^c\chi_3 | ^g\chi_2 \rangle$ & 0.479\\
$\langle ^c\chi_3 | ^g\chi_3 \rangle$ & 0.455\\
$\langle ^c\chi_3 | ^g\chi_4 \rangle$ & 0.662\\
\hline    
$\langle ^c\chi_4 | ^g\chi_0 \rangle$ & 0.022\\
$\langle ^c\chi_4 | ^g\chi_1 \rangle$ & 0.109\\
$\langle ^c\chi_4 | ^g\chi_2 \rangle$ & 0.308\\
$\langle ^c\chi_4 | ^g\chi_3 \rangle$ & 0.467\\
$\langle ^c\chi_4 | ^g\chi_4 \rangle$ & 0.331\\
\hline    
$\langle ^c\chi_0 | ^f\chi_0 \rangle$ & 0.773\\
$\langle ^c\chi_0 | ^f\chi_1 \rangle$ & -0.589\\
$\langle ^c\chi_0 | ^f\chi_2 \rangle$ & 0.233\\
$\langle ^c\chi_0 | ^f\chi_3 \rangle$ & -0.029\\
$\langle ^c\chi_0 | ^f\chi_4 \rangle$ & -0.020\\
\hline    
$\langle ^c\chi_1 | ^f\chi_0 \rangle$ & 0.504\\
$\langle ^c\chi_1 | ^f\chi_1 \rangle$ & 0.379\\
$\langle ^c\chi_1 | ^f\chi_2 \rangle$ & -0.671\\
$\langle ^c\chi_1 | ^f\chi_3 \rangle$ & 0.381\\
$\langle ^c\chi_1 | ^f\chi_4 \rangle$ & -0.070\\
\hline    
$\langle ^c\chi_2 | ^f\chi_0 \rangle$ & 0.317\\
$\langle ^c\chi_2 | ^f\chi_1 \rangle$ & 0.463\\
$\langle ^c\chi_2 | ^f\chi_2 \rangle$ & 0.091\\
$\langle ^c\chi_2 | ^f\chi_3 \rangle$ & -0.639\\
$\langle ^c\chi_2 | ^f\chi_4 \rangle$ & 0.497\\
\hline    
$\langle ^c\chi_3 | ^f\chi_0 \rangle$ & 0.183\\
$\langle ^c\chi_3 | ^f\chi_1 \rangle$ & 0.403\\
$\langle ^c\chi_3 | ^f\chi_2 \rangle$ & 0.324\\
$\langle ^c\chi_3 | ^f\chi_3 \rangle$ & -0.103\\
$\langle ^c\chi_3 | ^f\chi_4 \rangle$ & -0.551\\
\hline    
$\langle ^c\chi_4 | ^f\chi_0 \rangle$ & 0.102\\
$\langle ^c\chi_4 | ^f\chi_1 \rangle$ & 0.284\\
$\langle ^c\chi_4 | ^f\chi_2 \rangle$ & 0.399\\
$\langle ^c\chi_4 | ^f\chi_3 \rangle$ & 0.158\\
$\langle ^c\chi_4 | ^f\chi_4 \rangle$ & -0.216\\
\hline    
%  \end{tabular}
\end{longtable}

\tabcolsep 10.0pt
%\begin{table}
\begin{longtable}{@{}l|c|ccc@{}}
%  \centering
  \caption{RIXS  emission  peak  positions  (eV)  and  intensities  of
    elastic and energy-loss  peaks computed using FCFs for
    the transitions between the ground state (g), core-excited state, and
    final valence excited state (f). We consider only 
    $\nu = 0,1,2$ vibrational levels of the ground state.
    $\sigma^{RIXS}_{elec}$  used  for  computing  $\sigma
    ^{RIXS}_{vib}$  corresponds to  the  computed  cross sections  for
    snapshot \#2 (0.0367  a.u., 0.0720 a.u., and 0.0534  a.u.  for the
    elastic $1\pi\rightarrow 1s_O$,  near-degenerate elastic $1\pi\rightarrow 1s_O$, and 
    $2\sigma\rightarrow 1s_O$ transitions,
    respectively).    \protect\label{tbl:OH:VibSplit}} \endfirsthead \endhead
%  \begin{tabular}{l|c|ccc}
\hline
Transition & Position & $\sigma^{RIXS}_{vib}$ & $\sigma^{RIXS}_{vib}$ & $\sigma^{RIXS}_{vib}$  \\
 & & Non-thermal & T = 1,000 K & T = 10,000 K \\
\hline
g(v=0) $\rightarrow$ c(v=0) $\rightarrow$ f(v=0)  &  521.1377  &  0.02655351  &  0.02655354  &  0.02655354  \\
g(v=0) $\rightarrow$ c(v=0) $\rightarrow$ f(v=1)  &  520.7394  &  0.01542118  &  0.01542120  &  0.01542120  \\
g(v=0) $\rightarrow$ c(v=0) $\rightarrow$ f(v=2)  &  520.3411  &  0.00241791  &  0.00241792  &  0.00241792  \\
g(v=0) $\rightarrow$ c(v=0) $\rightarrow$ f(v=3)  &  519.9428  &  0.00003623  &  0.00003623  &  0.00003623  \\
g(v=0) $\rightarrow$ c(v=0) $\rightarrow$ f(v=4)  &  519.5445  &  0.00001819  &  0.00001819  &  0.00001819  \\
g(v=0) $\rightarrow$ c(v=1) $\rightarrow$ f(v=0)  &  521.6810  &  0.00188528  &  0.00188528  &  0.00188528  \\
g(v=0) $\rightarrow$ c(v=1) $\rightarrow$ f(v=1)  &  521.2827  &  0.00106676  &  0.00106676  &  0.00106676  \\
g(v=0) $\rightarrow$ c(v=1) $\rightarrow$ f(v=2)  &  520.8844  &  0.00333616  &  0.00333616  &  0.00333616  \\
g(v=0) $\rightarrow$ c(v=1) $\rightarrow$ f(v=3)  &  520.4861  &  0.00107330  &  0.00107330  &  0.00107330  \\
g(v=0) $\rightarrow$ c(v=1) $\rightarrow$ f(v=4)  &  520.0878  &  0.00003593  &  0.00003593  &  0.00003593  \\
g(v=0) $\rightarrow$ c(v=2) $\rightarrow$ f(v=0)  &  522.2244  &  0.00013003  &  0.00013004  &  0.00013004  \\
g(v=0) $\rightarrow$ c(v=2) $\rightarrow$ f(v=1)  &  521.8260  &  0.00027794  &  0.00027794  &  0.00027794  \\
g(v=0) $\rightarrow$ c(v=2) $\rightarrow$ f(v=2)  &  521.4277  &  0.00001066  &  0.00001066  &  0.00001066  \\
g(v=0) $\rightarrow$ c(v=2) $\rightarrow$ f(v=3)  &  521.0294  &  0.00052950  &  0.00052951  &  0.00052951  \\
g(v=0) $\rightarrow$ c(v=2) $\rightarrow$ f(v=4)  &  520.6311  &  0.00032043  &  0.00032044  &  0.00032044  \\
g(v=0) $\rightarrow$ c(v=3) $\rightarrow$ f(v=0)  &  522.7677  &  0.00000630  &  0.00000630  &  0.00000630  \\
g(v=0) $\rightarrow$ c(v=3) $\rightarrow$ f(v=1)  &  522.3694  &  0.00003062  &  0.00003062  &  0.00003062  \\
g(v=0) $\rightarrow$ c(v=3) $\rightarrow$ f(v=2)  &  521.9711  &  0.00001974  &  0.00001974  &  0.00001974  \\
g(v=0) $\rightarrow$ c(v=3) $\rightarrow$ f(v=3)  &  521.5728  &  0.00000202  &  0.00000202  &  0.00000202  \\
g(v=0) $\rightarrow$ c(v=3) $\rightarrow$ f(v=4)  &  521.1745  &  0.00005723  &  0.00005723  &  0.00005723  \\
g(v=0) $\rightarrow$ c(v=4) $\rightarrow$ f(v=0)  &  523.3110  &  0.00000027  &  0.00000027  &  0.00000027  \\
g(v=0) $\rightarrow$ c(v=4) $\rightarrow$ f(v=1)  &  522.9127  &  0.00000212  &  0.00000212  &  0.00000212  \\
g(v=0) $\rightarrow$ c(v=4) $\rightarrow$ f(v=2)  &  522.5144  &  0.00000419  &  0.00000419  &  0.00000419  \\
g(v=0) $\rightarrow$ c(v=4) $\rightarrow$ f(v=3)  &  522.1161  &  0.00000066  &  0.00000066  &  0.00000066  \\
g(v=0) $\rightarrow$ c(v=4) $\rightarrow$ f(v=4)  &  521.7178  &  0.00000122  &  0.00000122  &  0.00000122  \\
g(v=1) $\rightarrow$ c(v=0) $\rightarrow$ f(v=0)  &  521.1377  &  0.00514255  &  0.00002258  &  0.00298808  \\
g(v=1) $\rightarrow$ c(v=0) $\rightarrow$ f(v=1)  &  520.7394  &  0.00298658  &  0.00001311  &  0.00173535  \\
g(v=1) $\rightarrow$ c(v=0) $\rightarrow$ f(v=2)  &  520.3411  &  0.00046827  &  0.00000206  &  0.00027209  \\
g(v=1) $\rightarrow$ c(v=0) $\rightarrow$ f(v=3)  &  519.9428  &  0.00000702  &  0.00000003  &  0.00000408  \\
g(v=1) $\rightarrow$ c(v=0) $\rightarrow$ f(v=4)  &  519.5445  &  0.00000352  &  0.00000002  &  0.00000205  \\
g(v=1) $\rightarrow$ c(v=1) $\rightarrow$ f(v=0)  &  521.6810  &  0.00755524  &  0.00003314  &  0.00438999  \\
g(v=1) $\rightarrow$ c(v=1) $\rightarrow$ f(v=1)  &  521.2827  &  0.00427506  &  0.00001875  &  0.00248403  \\
g(v=1) $\rightarrow$ c(v=1) $\rightarrow$ f(v=2)  &  520.8844  &  0.01336963  &  0.00005864  &  0.00776845  \\
g(v=1) $\rightarrow$ c(v=1) $\rightarrow$ f(v=3)  &  520.4861  &  0.00430124  &  0.00001887  &  0.00249925  \\
g(v=1) $\rightarrow$ c(v=1) $\rightarrow$ f(v=4)  &  520.0878  &  0.00014398  &  0.00000063  &  0.00008366  \\
g(v=1) $\rightarrow$ c(v=2) $\rightarrow$ f(v=0)  &  522.2244  &  0.00111914  &  0.00000491  &  0.00065028  \\
g(v=1) $\rightarrow$ c(v=2) $\rightarrow$ f(v=1)  &  521.8260  &  0.00239209  &  0.00001049  &  0.00138992  \\
g(v=1) $\rightarrow$ c(v=2) $\rightarrow$ f(v=2)  &  521.4277  &  0.00009171  &  0.00000040  &  0.00005329  \\
g(v=1) $\rightarrow$ c(v=2) $\rightarrow$ f(v=3)  &  521.0294  &  0.00455720  &  0.00001999  &  0.00264796  \\
g(v=1) $\rightarrow$ c(v=2) $\rightarrow$ f(v=4)  &  520.6311  &  0.00275782  &  0.00001210  &  0.00160243  \\
g(v=1) $\rightarrow$ c(v=3) $\rightarrow$ f(v=0)  &  522.7677  &  0.00010521  &  0.00000046  &  0.00006113  \\
g(v=1) $\rightarrow$ c(v=3) $\rightarrow$ f(v=1)  &  522.3694  &  0.00051137  &  0.00000224  &  0.00029713  \\
g(v=1) $\rightarrow$ c(v=3) $\rightarrow$ f(v=2)  &  521.9711  &  0.00032963  &  0.00000145  &  0.00019153  \\
g(v=1) $\rightarrow$ c(v=3) $\rightarrow$ f(v=3)  &  521.5728  &  0.00003367  &  0.00000015  &  0.00001956  \\
g(v=1) $\rightarrow$ c(v=3) $\rightarrow$ f(v=4)  &  521.1745  &  0.00095564  &  0.00000420  &  0.00055528  \\
g(v=1) $\rightarrow$ c(v=4) $\rightarrow$ f(v=0)  &  523.3110  &  0.00000656  &  0.00000003  &  0.00000381  \\
g(v=1) $\rightarrow$ c(v=4) $\rightarrow$ f(v=1)  &  522.9127  &  0.00005078  &  0.00000022  &  0.00002950  \\
g(v=1) $\rightarrow$ c(v=4) $\rightarrow$ f(v=2)  &  522.5144  &  0.00010064  &  0.00000044  &  0.00005848  \\
g(v=1) $\rightarrow$ c(v=4) $\rightarrow$ f(v=3)  &  522.1161  &  0.00001573  &  0.00000007  &  0.00000914  \\
g(v=1) $\rightarrow$ c(v=4) $\rightarrow$ f(v=4)  &  521.7178  &  0.00002934  &  0.00000013  &  0.00001705  \\
g(v=2) $\rightarrow$ c(v=0) $\rightarrow$ f(v=0)  &  521.1377  &  0.00018808  &  0.00000000  &  0.00006349  \\
g(v=2) $\rightarrow$ c(v=0) $\rightarrow$ f(v=1)  &  520.7394  &  0.00010923  &  0.00000000  &  0.00003687  \\
g(v=2) $\rightarrow$ c(v=0) $\rightarrow$ f(v=2)  &  520.3411  &  0.00001713  &  0.00000000  &  0.00000578  \\
g(v=2) $\rightarrow$ c(v=0) $\rightarrow$ f(v=3)  &  519.9428  &  0.00000026  &  0.00000000  &  0.00000009  \\
g(v=2) $\rightarrow$ c(v=0) $\rightarrow$ f(v=4)  &  519.5445  &  0.00000013  &  0.00000000  &  0.00000004  \\
g(v=2) $\rightarrow$ c(v=1) $\rightarrow$ f(v=0)  &  521.6810  &  0.00388553  &  0.00000008  &  0.00131184  \\
g(v=2) $\rightarrow$ c(v=1) $\rightarrow$ f(v=1)  &  521.2827  &  0.00219859  &  0.00000005  &  0.00074229  \\
g(v=2) $\rightarrow$ c(v=1) $\rightarrow$ f(v=2)  &  520.8844  &  0.00687577  &  0.00000014  &  0.00232140  \\
g(v=2) $\rightarrow$ c(v=1) $\rightarrow$ f(v=3)  &  520.4861  &  0.00221206  &  0.00000005  &  0.00074684  \\
g(v=2) $\rightarrow$ c(v=1) $\rightarrow$ f(v=4)  &  520.0878  &  0.00007405  &  0.00000000  &  0.00002500  \\
g(v=2) $\rightarrow$ c(v=2) $\rightarrow$ f(v=0)  &  522.2244  &  0.00188759  &  0.00000004  &  0.00063729  \\
g(v=2) $\rightarrow$ c(v=2) $\rightarrow$ f(v=1)  &  521.8260  &  0.00403460  &  0.00000008  &  0.00136217  \\
g(v=2) $\rightarrow$ c(v=2) $\rightarrow$ f(v=2)  &  521.4277  &  0.00015468  &  0.00000000  &  0.00005222  \\
g(v=2) $\rightarrow$ c(v=2) $\rightarrow$ f(v=3)  &  521.0294  &  0.00768636  &  0.00000015  &  0.00259509  \\
g(v=2) $\rightarrow$ c(v=2) $\rightarrow$ f(v=4)  &  520.6311  &  0.00465145  &  0.00000009  &  0.00157043  \\
g(v=2) $\rightarrow$ c(v=3) $\rightarrow$ f(v=0)  &  522.7677  &  0.00040986  &  0.00000001  &  0.00013838  \\
g(v=2) $\rightarrow$ c(v=3) $\rightarrow$ f(v=1)  &  522.3694  &  0.00199210  &  0.00000003  &  0.00067257  \\
g(v=2) $\rightarrow$ c(v=3) $\rightarrow$ f(v=2)  &  521.9711  &  0.00128411  &  0.00000002  &  0.00043354  \\
g(v=2) $\rightarrow$ c(v=3) $\rightarrow$ f(v=3)  &  521.5728  &  0.00013116  &  0.00000000  &  0.00004428  \\
g(v=2) $\rightarrow$ c(v=3) $\rightarrow$ f(v=4)  &  521.1745  &  0.00372286  &  0.00000006  &  0.00125691  \\
g(v=2) $\rightarrow$ c(v=4) $\rightarrow$ f(v=0)  &  523.3110  &  0.00005241  &  0.00000000  &  0.00001769  \\
g(v=2) $\rightarrow$ c(v=4) $\rightarrow$ f(v=1)  &  522.9127  &  0.00040587  &  0.00000001  &  0.00013703  \\
g(v=2) $\rightarrow$ c(v=4) $\rightarrow$ f(v=2)  &  522.5144  &  0.00080442  &  0.00000002  &  0.00027159  \\
g(v=2) $\rightarrow$ c(v=4) $\rightarrow$ f(v=3)  &  522.1161  &  0.00012575  &  0.00000000  &  0.00004246  \\
g(v=2) $\rightarrow$ c(v=4) $\rightarrow$ f(v=4)  &  521.7178  &  0.00023453  &  0.00000000  &  0.00007918  \\
\hline
g(v=0) $\rightarrow$ c(v=0) $\rightarrow$ g(v=0)  &  525.0000  &  0.02546210  &  0.02546212  &  0.02546212  \\
g(v=0) $\rightarrow$ c(v=0) $\rightarrow$ g(v=1)  &  524.5322  &  0.00493118  &  0.00493118  &  0.00493118  \\
g(v=0) $\rightarrow$ c(v=0) $\rightarrow$ g(v=2)  &  524.0643  &  0.00018035  &  0.00018035  &  0.00018035  \\
g(v=0) $\rightarrow$ c(v=0) $\rightarrow$ g(v=3)  &  523.5965  &  0.00000075  &  0.00000075  &  0.00000075  \\
g(v=0) $\rightarrow$ c(v=0) $\rightarrow$ g(v=4)  &  523.1286  &  0.00000112  &  0.00000112  &  0.00000112  \\
g(v=0) $\rightarrow$ c(v=1) $\rightarrow$ g(v=0)  &  525.5433  &  0.00070810  &  0.00070810  &  0.00070810  \\
g(v=0) $\rightarrow$ c(v=1) $\rightarrow$ g(v=1)  &  525.0755  &  0.00283772  &  0.00283772  &  0.00283772  \\
g(v=0) $\rightarrow$ c(v=1) $\rightarrow$ g(v=2)  &  524.6076  &  0.00145939  &  0.00145939  &  0.00145939  \\
g(v=0) $\rightarrow$ c(v=1) $\rightarrow$ g(v=3)  &  524.1398  &  0.00009249  &  0.00009249  &  0.00009249  \\
g(v=0) $\rightarrow$ c(v=1) $\rightarrow$ g(v=4)  &  523.6719  &  0.00000028  &  0.00000028  &  0.00000028  \\
g(v=0) $\rightarrow$ c(v=2) $\rightarrow$ g(v=0)  &  526.0867  &  0.00002162  &  0.00002162  &  0.00002162  \\
g(v=0) $\rightarrow$ c(v=2) $\rightarrow$ g(v=1)  &  525.6188  &  0.00018607  &  0.00018607  &  0.00018607  \\
g(v=0) $\rightarrow$ c(v=2) $\rightarrow$ g(v=2)  &  525.1510  &  0.00031383  &  0.00031383  &  0.00031383  \\
g(v=0) $\rightarrow$ c(v=2) $\rightarrow$ g(v=3)  &  524.6831  &  0.00033615  &  0.00033616  &  0.00033616  \\
g(v=0) $\rightarrow$ c(v=2) $\rightarrow$ g(v=4)  &  524.2153  &  0.00003276  &  0.00003276  &  0.00003276  \\
g(v=0) $\rightarrow$ c(v=3) $\rightarrow$ g(v=0)  &  526.6300  &  0.00000046  &  0.00000046  &  0.00000046  \\
g(v=0) $\rightarrow$ c(v=3) $\rightarrow$ g(v=1)  &  526.1621  &  0.00000765  &  0.00000765  &  0.00000765  \\
g(v=0) $\rightarrow$ c(v=3) $\rightarrow$ g(v=2)  &  525.6943  &  0.00002981  &  0.00002981  &  0.00002981  \\
g(v=0) $\rightarrow$ c(v=3) $\rightarrow$ g(v=3)  &  525.2264  &  0.00002688  &  0.00002688  &  0.00002688  \\
g(v=0) $\rightarrow$ c(v=3) $\rightarrow$ g(v=4)  &  524.7586  &  0.00005678  &  0.00005678  &  0.00005678  \\
g(v=0) $\rightarrow$ c(v=4) $\rightarrow$ g(v=0)  &  527.1733  &  0.00000001  &  0.00000001  &  0.00000001  \\
g(v=0) $\rightarrow$ c(v=4) $\rightarrow$ g(v=1)  &  526.7055  &  0.00000021  &  0.00000021  &  0.00000021  \\
g(v=0) $\rightarrow$ c(v=4) $\rightarrow$ g(v=2)  &  526.2376  &  0.00000171  &  0.00000171  &  0.00000171  \\
g(v=0) $\rightarrow$ c(v=4) $\rightarrow$ g(v=3)  &  525.7698  &  0.00000395  &  0.00000395  &  0.00000395  \\
g(v=0) $\rightarrow$ c(v=4) $\rightarrow$ g(v=4)  &  525.3019  &  0.00000199  &  0.00000199  &  0.00000199  \\
g(v=1) $\rightarrow$ c(v=0) $\rightarrow$ g(v=0)  &  525.0000  &  0.00493118  &  0.00002165  &  0.00286526  \\
g(v=1) $\rightarrow$ c(v=0) $\rightarrow$ g(v=1)  &  524.5322  &  0.00095501  &  0.00000419  &  0.00055491  \\
g(v=1) $\rightarrow$ c(v=0) $\rightarrow$ g(v=2)  &  524.0643  &  0.00003493  &  0.00000015  &  0.00002030  \\
g(v=1) $\rightarrow$ c(v=0) $\rightarrow$ g(v=3)  &  523.5965  &  0.00000015  &  0.00000000  &  0.00000008  \\
g(v=1) $\rightarrow$ c(v=0) $\rightarrow$ g(v=4)  &  523.1286  &  0.00000022  &  0.00000000  &  0.00000013  \\
g(v=1) $\rightarrow$ c(v=1) $\rightarrow$ g(v=0)  &  525.5433  &  0.00283772  &  0.00001245  &  0.00164886  \\
g(v=1) $\rightarrow$ c(v=1) $\rightarrow$ g(v=1)  &  525.0755  &  0.01137214  &  0.00004988  &  0.00660781  \\
g(v=1) $\rightarrow$ c(v=1) $\rightarrow$ g(v=2)  &  524.6076  &  0.00584850  &  0.00002565  &  0.00339828  \\
g(v=1) $\rightarrow$ c(v=1) $\rightarrow$ g(v=3)  &  524.1398  &  0.00037065  &  0.00000163  &  0.00021537  \\
g(v=1) $\rightarrow$ c(v=1) $\rightarrow$ g(v=4)  &  523.6719  &  0.00000113  &  0.00000000  &  0.00000065  \\
g(v=1) $\rightarrow$ c(v=2) $\rightarrow$ g(v=0)  &  526.0867  &  0.00018607  &  0.00000082  &  0.00010811  \\
g(v=1) $\rightarrow$ c(v=2) $\rightarrow$ g(v=1)  &  525.6188  &  0.00160140  &  0.00000702  &  0.00093049  \\
g(v=1) $\rightarrow$ c(v=2) $\rightarrow$ g(v=2)  &  525.1510  &  0.00270099  &  0.00001185  &  0.00156941  \\
g(v=1) $\rightarrow$ c(v=2) $\rightarrow$ g(v=3)  &  524.6831  &  0.00289312  &  0.00001269  &  0.00168104  \\
g(v=1) $\rightarrow$ c(v=2) $\rightarrow$ g(v=4)  &  524.2153  &  0.00028196  &  0.00000124  &  0.00016383  \\
g(v=1) $\rightarrow$ c(v=3) $\rightarrow$ g(v=0)  &  526.6300  &  0.00000765  &  0.00000003  &  0.00000445  \\
g(v=1) $\rightarrow$ c(v=3) $\rightarrow$ g(v=1)  &  526.1621  &  0.00012781  &  0.00000056  &  0.00007426  \\
g(v=1) $\rightarrow$ c(v=3) $\rightarrow$ g(v=2)  &  525.6943  &  0.00049789  &  0.00000219  &  0.00028930  \\
g(v=1) $\rightarrow$ c(v=3) $\rightarrow$ g(v=3)  &  525.2264  &  0.00044892  &  0.00000197  &  0.00026085  \\
g(v=1) $\rightarrow$ c(v=3) $\rightarrow$ g(v=4)  &  524.7586  &  0.00094828  &  0.00000416  &  0.00055100  \\
g(v=1) $\rightarrow$ c(v=4) $\rightarrow$ g(v=0)  &  527.1733  &  0.00000021  &  0.00000000  &  0.00000012  \\
g(v=1) $\rightarrow$ c(v=4) $\rightarrow$ g(v=1)  &  526.7055  &  0.00000514  &  0.00000002  &  0.00000299  \\
g(v=1) $\rightarrow$ c(v=4) $\rightarrow$ g(v=2)  &  526.2376  &  0.00004108  &  0.00000018  &  0.00002387  \\
g(v=1) $\rightarrow$ c(v=4) $\rightarrow$ g(v=3)  &  525.7698  &  0.00009485  &  0.00000042  &  0.00005511  \\
g(v=1) $\rightarrow$ c(v=4) $\rightarrow$ g(v=4)  &  525.3019  &  0.00004767  &  0.00000021  &  0.00002770  \\
g(v=2) $\rightarrow$ c(v=0) $\rightarrow$ g(v=0)  &  525.0000  &  0.00018035  &  0.00000000  &  0.00006088  \\
g(v=2) $\rightarrow$ c(v=0) $\rightarrow$ g(v=1)  &  524.5322  &  0.00003493  &  0.00000000  &  0.00001179  \\
g(v=2) $\rightarrow$ c(v=0) $\rightarrow$ g(v=2)  &  524.0643  &  0.00000128  &  0.00000000  &  0.00000043  \\
g(v=2) $\rightarrow$ c(v=0) $\rightarrow$ g(v=3)  &  523.5965  &  0.00000001  &  0.00000000  &  0.00000000  \\
g(v=2) $\rightarrow$ c(v=0) $\rightarrow$ g(v=4)  &  523.1286  &  0.00000001  &  0.00000000  &  0.00000000  \\
g(v=2) $\rightarrow$ c(v=1) $\rightarrow$ g(v=0)  &  525.5433  &  0.00145939  &  0.00000003  &  0.00049272  \\
g(v=2) $\rightarrow$ c(v=1) $\rightarrow$ g(v=1)  &  525.0755  &  0.00584850  &  0.00000012  &  0.00197457  \\
g(v=2) $\rightarrow$ c(v=1) $\rightarrow$ g(v=2)  &  524.6076  &  0.00300778  &  0.00000006  &  0.00101549  \\
g(v=2) $\rightarrow$ c(v=1) $\rightarrow$ g(v=3)  &  524.1398  &  0.00019062  &  0.00000000  &  0.00006436  \\
g(v=2) $\rightarrow$ c(v=1) $\rightarrow$ g(v=4)  &  523.6719  &  0.00000058  &  0.00000000  &  0.00000020  \\
g(v=2) $\rightarrow$ c(v=2) $\rightarrow$ g(v=0)  &  526.0867  &  0.00031383  &  0.00000001  &  0.00010595  \\
g(v=2) $\rightarrow$ c(v=2) $\rightarrow$ g(v=1)  &  525.6188  &  0.00270099  &  0.00000005  &  0.00091191  \\
g(v=2) $\rightarrow$ c(v=2) $\rightarrow$ g(v=2)  &  525.1510  &  0.00455560  &  0.00000009  &  0.00153807  \\
g(v=2) $\rightarrow$ c(v=2) $\rightarrow$ g(v=3)  &  524.6831  &  0.00487965  &  0.00000010  &  0.00164748  \\
g(v=2) $\rightarrow$ c(v=2) $\rightarrow$ g(v=4)  &  524.2153  &  0.00047557  &  0.00000001  &  0.00016056  \\
g(v=2) $\rightarrow$ c(v=3) $\rightarrow$ g(v=0)  &  526.6300  &  0.00002981  &  0.00000000  &  0.00001007  \\
g(v=2) $\rightarrow$ c(v=3) $\rightarrow$ g(v=1)  &  526.1621  &  0.00049789  &  0.00000001  &  0.00016810  \\
g(v=2) $\rightarrow$ c(v=3) $\rightarrow$ g(v=2)  &  525.6943  &  0.00193960  &  0.00000003  &  0.00065485  \\
g(v=2) $\rightarrow$ c(v=3) $\rightarrow$ g(v=3)  &  525.2264  &  0.00174885  &  0.00000003  &  0.00059045  \\
g(v=2) $\rightarrow$ c(v=3) $\rightarrow$ g(v=4)  &  524.7586  &  0.00369416  &  0.00000006  &  0.00124722  \\
g(v=2) $\rightarrow$ c(v=4) $\rightarrow$ g(v=0)  &  527.1733  &  0.00000171  &  0.00000000  &  0.00000058  \\
g(v=2) $\rightarrow$ c(v=4) $\rightarrow$ g(v=1)  &  526.7055  &  0.00004108  &  0.00000000  &  0.00001387  \\
g(v=2) $\rightarrow$ c(v=4) $\rightarrow$ g(v=2)  &  526.2376  &  0.00032841  &  0.00000001  &  0.00011088  \\
g(v=2) $\rightarrow$ c(v=4) $\rightarrow$ g(v=3)  &  525.7698  &  0.00075818  &  0.00000002  &  0.00025598  \\
g(v=2) $\rightarrow$ c(v=4) $\rightarrow$ g(v=4)  &  525.3019  &  0.00038105  &  0.00000001  &  0.00012865  \\
\hline
g(v=0) $\rightarrow$ c(v=0) $\rightarrow$ dg(v=0)  &  525.0000  &  0.04993764  &  0.04993768  &  0.04993768  \\
g(v=0) $\rightarrow$ c(v=0) $\rightarrow$ dg(v=1)  &  524.5322  &  0.00967129  &  0.00967130  &  0.00967130  \\
g(v=0) $\rightarrow$ c(v=0) $\rightarrow$ dg(v=2)  &  524.0643  &  0.00035371  &  0.00035371  &  0.00035371  \\
g(v=0) $\rightarrow$ c(v=0) $\rightarrow$ dg(v=3)  &  523.5965  &  0.00000148  &  0.00000148  &  0.00000148  \\
g(v=0) $\rightarrow$ c(v=0) $\rightarrow$ dg(v=4)  &  523.1286  &  0.00000220  &  0.00000220  &  0.00000220  \\
g(v=0) $\rightarrow$ c(v=1) $\rightarrow$ dg(v=0)  &  525.5433  &  0.00138877  &  0.00138877  &  0.00138877  \\
g(v=0) $\rightarrow$ c(v=1) $\rightarrow$ dg(v=1)  &  525.0755  &  0.00556548  &  0.00556548  &  0.00556548  \\
g(v=0) $\rightarrow$ c(v=1) $\rightarrow$ dg(v=2)  &  524.6076  &  0.00286223  &  0.00286223  &  0.00286223  \\
g(v=0) $\rightarrow$ c(v=1) $\rightarrow$ dg(v=3)  &  524.1398  &  0.00018139  &  0.00018139  &  0.00018139  \\
g(v=0) $\rightarrow$ c(v=1) $\rightarrow$ dg(v=4)  &  523.6719  &  0.00000055  &  0.00000055  &  0.00000055  \\
g(v=0) $\rightarrow$ c(v=2) $\rightarrow$ dg(v=0)  &  526.0867  &  0.00004240  &  0.00004240  &  0.00004240  \\
g(v=0) $\rightarrow$ c(v=2) $\rightarrow$ dg(v=1)  &  525.6188  &  0.00036492  &  0.00036493  &  0.00036493  \\
g(v=0) $\rightarrow$ c(v=2) $\rightarrow$ dg(v=2)  &  525.1510  &  0.00061549  &  0.00061551  &  0.00061551  \\
g(v=0) $\rightarrow$ c(v=2) $\rightarrow$ dg(v=3)  &  524.6831  &  0.00065928  &  0.00065929  &  0.00065929  \\
g(v=0) $\rightarrow$ c(v=2) $\rightarrow$ dg(v=4)  &  524.2153  &  0.00006425  &  0.00006425  &  0.00006425  \\
g(v=0) $\rightarrow$ c(v=3) $\rightarrow$ dg(v=0)  &  526.6300  &  0.00000090  &  0.00000090  &  0.00000090  \\
g(v=0) $\rightarrow$ c(v=3) $\rightarrow$ dg(v=1)  &  526.1621  &  0.00001501  &  0.00001501  &  0.00001501  \\
g(v=0) $\rightarrow$ c(v=3) $\rightarrow$ dg(v=2)  &  525.6943  &  0.00005847  &  0.00005847  &  0.00005847  \\
g(v=0) $\rightarrow$ c(v=3) $\rightarrow$ dg(v=3)  &  525.2264  &  0.00005272  &  0.00005272  &  0.00005272  \\
g(v=0) $\rightarrow$ c(v=3) $\rightarrow$ dg(v=4)  &  524.7586  &  0.00011137  &  0.00011137  &  0.00011137  \\
g(v=0) $\rightarrow$ c(v=4) $\rightarrow$ dg(v=0)  &  527.1733  &  0.00000002  &  0.00000002  &  0.00000002  \\
g(v=0) $\rightarrow$ c(v=4) $\rightarrow$ dg(v=1)  &  526.7055  &  0.00000042  &  0.00000042  &  0.00000042  \\
g(v=0) $\rightarrow$ c(v=4) $\rightarrow$ dg(v=2)  &  526.2376  &  0.00000336  &  0.00000336  &  0.00000336  \\
g(v=0) $\rightarrow$ c(v=4) $\rightarrow$ dg(v=3)  &  525.7698  &  0.00000775  &  0.00000775  &  0.00000775  \\
g(v=0) $\rightarrow$ c(v=4) $\rightarrow$ dg(v=4)  &  525.3019  &  0.00000390  &  0.00000390  &  0.00000390  \\
g(v=1) $\rightarrow$ c(v=0) $\rightarrow$ dg(v=0)  &  525.0000  &  0.00967129  &  0.00004246  &  0.00561951  \\
g(v=1) $\rightarrow$ c(v=0) $\rightarrow$ dg(v=1)  &  524.5322  &  0.00187301  &  0.00000822  &  0.00108832  \\
g(v=1) $\rightarrow$ c(v=0) $\rightarrow$ dg(v=2)  &  524.0643  &  0.00006850  &  0.00000030  &  0.00003980  \\
g(v=1) $\rightarrow$ c(v=0) $\rightarrow$ dg(v=3)  &  523.5965  &  0.00000029  &  0.00000000  &  0.00000017  \\
g(v=1) $\rightarrow$ c(v=0) $\rightarrow$ dg(v=4)  &  523.1286  &  0.00000043  &  0.00000000  &  0.00000025  \\
g(v=1) $\rightarrow$ c(v=1) $\rightarrow$ dg(v=0)  &  525.5433  &  0.00556548  &  0.00002441  &  0.00323384  \\
g(v=1) $\rightarrow$ c(v=1) $\rightarrow$ dg(v=1)  &  525.0755  &  0.02230365  &  0.00009782  &  0.01295958  \\
g(v=1) $\rightarrow$ c(v=1) $\rightarrow$ dg(v=2)  &  524.6076  &  0.01147039  &  0.00005031  &  0.00666490  \\
g(v=1) $\rightarrow$ c(v=1) $\rightarrow$ dg(v=3)  &  524.1398  &  0.00072693  &  0.00000319  &  0.00042239  \\
g(v=1) $\rightarrow$ c(v=1) $\rightarrow$ dg(v=4)  &  523.6719  &  0.00000221  &  0.00000001  &  0.00000128  \\
g(v=1) $\rightarrow$ c(v=2) $\rightarrow$ dg(v=0)  &  526.0867  &  0.00036492  &  0.00000160  &  0.00021204  \\
g(v=1) $\rightarrow$ c(v=2) $\rightarrow$ dg(v=1)  &  525.6188  &  0.00314075  &  0.00001378  &  0.00182493  \\
g(v=1) $\rightarrow$ c(v=2) $\rightarrow$ dg(v=2)  &  525.1510  &  0.00529733  &  0.00002323  &  0.00307801  \\
g(v=1) $\rightarrow$ c(v=2) $\rightarrow$ dg(v=3)  &  524.6831  &  0.00567414  &  0.00002489  &  0.00329696  \\
g(v=1) $\rightarrow$ c(v=2) $\rightarrow$ dg(v=4)  &  524.2153  &  0.00055300  &  0.00000243  &  0.00032132  \\
g(v=1) $\rightarrow$ c(v=3) $\rightarrow$ dg(v=0)  &  526.6300  &  0.00001501  &  0.00000007  &  0.00000872  \\
g(v=1) $\rightarrow$ c(v=3) $\rightarrow$ dg(v=1)  &  526.1621  &  0.00025066  &  0.00000110  &  0.00014565  \\
g(v=1) $\rightarrow$ c(v=3) $\rightarrow$ dg(v=2)  &  525.6943  &  0.00097648  &  0.00000429  &  0.00056739  \\
g(v=1) $\rightarrow$ c(v=3) $\rightarrow$ dg(v=3)  &  525.2264  &  0.00088045  &  0.00000387  &  0.00051159  \\
g(v=1) $\rightarrow$ c(v=3) $\rightarrow$ dg(v=4)  &  524.7586  &  0.00185981  &  0.00000816  &  0.00108066  \\
g(v=1) $\rightarrow$ c(v=4) $\rightarrow$ dg(v=0)  &  527.1733  &  0.00000042  &  0.00000000  &  0.00000024  \\
g(v=1) $\rightarrow$ c(v=4) $\rightarrow$ dg(v=1)  &  526.7055  &  0.00001008  &  0.00000004  &  0.00000586  \\
g(v=1) $\rightarrow$ c(v=4) $\rightarrow$ dg(v=2)  &  526.2376  &  0.00008058  &  0.00000035  &  0.00004682  \\
g(v=1) $\rightarrow$ c(v=4) $\rightarrow$ dg(v=3)  &  525.7698  &  0.00018603  &  0.00000082  &  0.00010809  \\
g(v=1) $\rightarrow$ c(v=4) $\rightarrow$ dg(v=4)  &  525.3019  &  0.00009350  &  0.00000041  &  0.00005433  \\
g(v=2) $\rightarrow$ c(v=0) $\rightarrow$ dg(v=0)  &  525.0000  &  0.00035371  &  0.00000000  &  0.00011939  \\
g(v=2) $\rightarrow$ c(v=0) $\rightarrow$ dg(v=1)  &  524.5322  &  0.00006850  &  0.00000000  &  0.00002312  \\
g(v=2) $\rightarrow$ c(v=0) $\rightarrow$ dg(v=2)  &  524.0643  &  0.00000251  &  0.00000000  &  0.00000085  \\
g(v=2) $\rightarrow$ c(v=0) $\rightarrow$ dg(v=3)  &  523.5965  &  0.00000001  &  0.00000000  &  0.00000000  \\
g(v=2) $\rightarrow$ c(v=0) $\rightarrow$ dg(v=4)  &  523.1286  &  0.00000002  &  0.00000000  &  0.00000001  \\
g(v=2) $\rightarrow$ c(v=1) $\rightarrow$ dg(v=0)  &  525.5433  &  0.00286223  &  0.00000006  &  0.00096635  \\
g(v=2) $\rightarrow$ c(v=1) $\rightarrow$ dg(v=1)  &  525.0755  &  0.01147039  &  0.00000024  &  0.00387264  \\
g(v=2) $\rightarrow$ c(v=1) $\rightarrow$ dg(v=2)  &  524.6076  &  0.00589903  &  0.00000012  &  0.00199163  \\
g(v=2) $\rightarrow$ c(v=1) $\rightarrow$ dg(v=3)  &  524.1398  &  0.00037385  &  0.00000001  &  0.00012622  \\
g(v=2) $\rightarrow$ c(v=1) $\rightarrow$ dg(v=4)  &  523.6719  &  0.00000114  &  0.00000000  &  0.00000038  \\
g(v=2) $\rightarrow$ c(v=2) $\rightarrow$ dg(v=0)  &  526.0867  &  0.00061549  &  0.00000001  &  0.00020780  \\
g(v=2) $\rightarrow$ c(v=2) $\rightarrow$ dg(v=1)  &  525.6188  &  0.00529733  &  0.00000011  &  0.00178849  \\
g(v=2) $\rightarrow$ c(v=2) $\rightarrow$ dg(v=2)  &  525.1510  &  0.00893469  &  0.00000018  &  0.00301655  \\
g(v=2) $\rightarrow$ c(v=2) $\rightarrow$ dg(v=3)  &  524.6831  &  0.00957024  &  0.00000019  &  0.00323112  \\
g(v=2) $\rightarrow$ c(v=2) $\rightarrow$ dg(v=4)  &  524.2153  &  0.00093272  &  0.00000002  &  0.00031491  \\
g(v=2) $\rightarrow$ c(v=3) $\rightarrow$ dg(v=0)  &  526.6300  &  0.00005847  &  0.00000000  &  0.00001974  \\
g(v=2) $\rightarrow$ c(v=3) $\rightarrow$ dg(v=1)  &  526.1621  &  0.00097648  &  0.00000002  &  0.00032968  \\
g(v=2) $\rightarrow$ c(v=3) $\rightarrow$ dg(v=2)  &  525.6943  &  0.00380404  &  0.00000007  &  0.00128432  \\
g(v=2) $\rightarrow$ c(v=3) $\rightarrow$ dg(v=3)  &  525.2264  &  0.00342994  &  0.00000006  &  0.00115801  \\
g(v=2) $\rightarrow$ c(v=3) $\rightarrow$ dg(v=4)  &  524.7586  &  0.00724519  &  0.00000013  &  0.00244612  \\
g(v=2) $\rightarrow$ c(v=4) $\rightarrow$ dg(v=0)  &  527.1733  &  0.00000336  &  0.00000000  &  0.00000113  \\
g(v=2) $\rightarrow$ c(v=4) $\rightarrow$ dg(v=1)  &  526.7055  &  0.00008058  &  0.00000000  &  0.00002721  \\
g(v=2) $\rightarrow$ c(v=4) $\rightarrow$ dg(v=2)  &  526.2376  &  0.00064409  &  0.00000001  &  0.00021746  \\
g(v=2) $\rightarrow$ c(v=4) $\rightarrow$ dg(v=3)  &  525.7698  &  0.00148699  &  0.00000003  &  0.00050204  \\
g(v=2) $\rightarrow$ c(v=4) $\rightarrow$ dg(v=4)  &  525.3019  &  0.00074734  &  0.00000002  &  0.00025232  \\
\hline
%  \end{tabular}
\end{longtable}